\documentclass[apj]{emulateapj}

\usepackage{amssymb} 
\usepackage{amsmath} 
\usepackage{graphicx}
\usepackage{latexsym} 
\usepackage{dcolumn}
\usepackage{bm}
\input{epsf} 

\newcommand{\be}{\begin{equation}}
\newcommand{\ee}{\end{equation}} 
\newcommand{\ba}{\begin{eqnarray}}
\newcommand{\ea}{\end{eqnarray}} 
\newcommand{\no}{\nonumber}
\newcommand{\bfi}{\begin{figure} \epsfxsize=8cm \epsffile}
\newcommand{\bfig}{\begin{figure*} \epsfxsize=15cm \epsffile}
\newcommand{\efi}{\end{figure}} 
\newcommand{\efig}{\end{figure*}}
\newcommand{\bi}{\begin{itemize}} 
\newcommand{\ei}{\end{itemize}}
\newcommand{\mpch}{h^{-1} {\rm Mpc}} 
\newcommand{\hmpc}{h {\rm Mpc}^{-1}}

\newcommand{\dif}{\mathrm{d}} 
\newcommand{\etal}{{et al. }}

\newcommand{\rmx}{{\times}}
\newcommand{\rmt}{{\rm t}} 
\newcommand{\obs}{{\rm obs}}

\slugcomment{Apj, 803, 46 (2015)}

\makeatletter

\shorttitle{Source-lens Clustering}
\shortauthors{Yu et al.}

\begin{document}

\title{The source-lens clustering effect in the context of lensing
  tomography and its self-calibration}

\author{Yu Yu\altaffilmark{1,*}, Pengjie Zhang\altaffilmark{2,1,$\dagger$},
  Weipeng Lin\altaffilmark{1,$\ddagger$},  and Weiguang Cui\altaffilmark{3,4}}
\altaffiltext{1}{Key laboratory for research in galaxies and cosmology,
Shanghai Astronomical Observatory, Chinese Academy of Science, 80
Nandan Road, Shanghai, China, 200030}
\altaffiltext{2}{Center for Astronomy and Astrophysics, Department of Physics and Astronomy,
Shanghai Jiao Tong University, 955 Jianchuan Road, Shanghai, 200240}
\altaffiltext{3}{ICRAR, University of Western Australia, 35 Stirling Highway,
Crawley, Western Australia 6009, Australia}
\altaffiltext{4}{ARC Centre of Excellence for All-Sky Astrophysics (CAASTRO)}
\altaffiltext{*}{yuyu22@shao.ac.cn}
\altaffiltext{$\dagger$}{zhangpj@sjtu.edu.cn}
\altaffiltext{$\ddagger$}{linwp@shao.ac.cn}
\begin{abstract}
Cosmic shear can only be measured where there are galaxies. 
This source-lens clustering (SLC) effect has two sources,
 intrinsic source clustering and cosmic magnification (magnification/size bias).
Lensing tomography can suppress the former. 
However, this reduction is limited by the existence of photo-z error and non-zero redshift bin width. 
Furthermore, the SLC induced by cosmic magnification can not be reduced by lensing tomography.   
Through N-body simulations, we quantify the impact of SLC on the lensing power spectrum in the context of lensing tomography.
We consider both the standard estimator and the pixel-based estimator.
We find that none of them can satisfactorily handle both sources of SLC.
 (1) For the standard estimator, the SLC induced by both sources
 can bias the lensing power spectrum by $\mathcal{O}(1\%)$-$\mathcal{O}(10\%)$.
Intrinsic source clustering also increases statistical uncertainties in the measured lensing power spectrum.   
However, the standard estimator suppresses the intrinsic source clustering in cross spectrum.
(2) In contrast, the pixel-based estimator suppresses the SLC by cosmic magnification.
However, it fails to suppress the SLC by intrinsic source clustering
 and the measured lensing power spectrum can be biased low by $\mathcal{O}(1\%)$-$\mathcal{O}(10\%)$.  
In a word, for typical photo-z error ($\sigma_z/(1+z)=0.05$) and photo-z bin size ($\Delta z^P=0.2$),  
SLC alters the lensing E-mode power spectrum by $1\%$-$10\%$, at $\ell\sim
10^3$ and $z_s\sim 1$ of particular interest to weak lensing cosmology.
Therefore the SLC is a severe systematic for cosmology in Stage-IV lensing surveys.   
We present useful scaling relations to self-calibrate the SLC effect.
\end{abstract}


\section{Introduction}
\label{sec:introduction}

Weak gravitational lensing --- a powerful probe of the dark universe
\citep{Bartelmann01,Refregier03,Munshi08,Hoekstra08,Bartelmann10,LSSTwhite}
--- still suffers from many systematic errors.  
Ongoing lensing surveys like the dark energy survey (DES) \footnote{Dark Energy Survey, http://www.darkenergysurvey.org} 
and upcoming surveys such as Euclid \footnote{Euclid, http://sci.esa.int/euclid/}, 
LSST \footnote{Large Synoptic Survey Telescope, http://www.lsst.org}, 
and Subaru-HSC \footnote{Subaru Hyper Suprime-Cam, http://www.naoj.org/Projects/HSC}
 will achieve $1\%$ level or better statistical precision.  
This puts stringent requirements on the accuracy of observational
measurement and theoretical modelling \citep{Huterer06}.

There are various known systematic errors $\sim 1\%$ level in weak
lensing measurement  (e.g. \citealt{LSSTwhite} and references therein), 
including point spread function, photometric redshift errors and
galaxy intrinsic alignment.  This motivates highly intensive and
comprehensive efforts to understand  and calibrate these systematic errors
(e.g. \citealt{Jing02,Hirata04,Jain06,STEP,Heymans06,Waerbeke06,Ma06,Bridle07,Joachimi08,GREAT08,
Okumura09a,Okumura09b,Joachimi10,Zhang10b,Zhang10c,Zhang10a,Bernstein10,Zhang11,
Troxel12a,Troxel12b,Troxel12c,Chang12,Chang13,Hamana13,GREAT3}). 
Meanwhile, there are various sources of systematic errors in theoretical calculation of weak lensing.  
(1) The standard treatment adopts the Born approximation.  
Hence two systematic errors, Born correction and lens-lens coupling
\citep{Schneider98,Dodelson05a,Dodelson05b,Dodelson06,Hilbert09,Becker13}, arise.  
It also approximates the measured  reduced shear $g\equiv\gamma/(1+\kappa)$ as $\gamma$ \citep{Schneider98,Dodelson05b,Dodelson06}.  
(2) The standard treatment neglects the baryon effect on the evolution of the matter density distribution.  
Studies \citep{White04,Zhan04,Jing06,Rudd08,Zentner08,vanDaalen11,Semboloni11,Semboloni13,Zentner13,vanDaalen14,Eifler2014,Joachim14} 
show that baryons 
can affect the weak lensing power spectrum by $\sim 10\%$ at scales of interest ($\ell\sim 10^3$).  
(3) Accurate non-linear power spectrum prediction relies on simulations.
The modeling at small scales $k\gtrsim 1\hmpc$ can have significant infuence on multipoles of $\ell\sim 10^3$.
Inproperly resolved dark matter substructures also hampers the accurate modeling of lensing signal at small scales (\cite{Hagan05,McDonald06}).
(4) It also ignores various selection biases such as the source-lens clustering (SLC).  
This paper aims to quantify SLC in the context of lensing tomography 
of typical photometric redshift bin size $\Delta z^P=0.2$.

The SLC effect \citep{Bernardeau98,Hamana01} arises from the fact that
we can only measure cosmic shear where there are (source) galaxies.
Any spatial correlation between the distribution of source galaxies and the lens field (source-lens clustering) 
can lead to bias in sampling the lensing field and hence lead to biased measurement of lensing statistics.  
There are two known sources causing such spatial correlation.  
(1) One is the overlapping of source redshift distribution and lens redshift distribution. 
Such overlapping is inevitable in reality.  
We need sufficiently wide redshift bin with sufficient amount of source galaxies to suppress shape measurement noise.  
For source galaxies in real redshift range of $z_1<z_s<z_2$, lenses distribute in the range $0<z_L<z_2$.  
So sources and lenses overlap at $(z_1,z_2)$.  
This overlap causes the intrinsic clustering of source galaxies ($\delta_g$) to be correlated with 
their cosmic shear $\gamma_{1,2}$ and convergence $\kappa$, for both tracing the underlying matter distribution.  
This correlation leads to biased sampling of the lensing field.  
Photometric redshift errors further broaden the galaxy distribution in redshift, making the situation worse.  
(2) The other is the magnification bias \citep{Hamana01,Romero07}, size
bias \citep{Schmidt09}  
or any other selection effects depending on the lensing field. 
For brevity we refer it as cosmic magnification.
Lensing changes the galaxy flux/size and hence alters the galaxy number density,  $\delta_g\rightarrow \delta_g+g\kappa$
at leading order.  
Here, $g$ is a function of galaxy flux and size and we refer it to magnification prefactor.
It ($g\kappa$) is perfectly correlated with the lensing field ($\kappa$).  
So this effect persists even for source redshift distribution of infinitesimal width.

Impacts of SLC on various weak lensing statistics have been investigated.
\cite{Bernardeau98} showed that SLC by the galaxy intrinsic clustering 
alters lensing skewness and kurtosis at leading order.  
\cite{Hamana01} extended the investigation of skewness 
to a wide variety of redshift distributions and bias parameters. 
\cite{Hamana02} studied SLC by the cosmic magnification using semi-analytical approach. 
They found that lensing magnification can affect the convergence skewness by $\la 3\%$.  
\cite{Schneider02} pointed out that intrinsic source clustering can produce B-mode 
and presented estimation based on analytical approach.
\cite{Romero07} presented a tool coupling N-body simulations 
to a semi-analytical model of galaxy formation to perform mock weak lensing measurements, 
and studied the intrinsic source clustering effect on the probability density distribution (PDF),
variance, skewness and kurtosis of the convergence.  
They found that $\sigma_8$ can be biased by $2\%$-$5\%$.  
\cite{Schmidt09} worked on the effect of magnification and size bias 
on the estimation of shear correlation function.  
They concluded that this lensing bias should be taken seriously.  
Their work extended the investigation to 2-point correlation function/power spectrum, compared to the previous works 
\citep{Hamana01,Hamana02} on 1-point variance and skewness.  
\cite{Valageas14} adopted hierarchical ansatz and semi-analytic model
to estimate the bias  
caused by the intrinsic source clustering on weak lensing 2-point and 3-point estimators.

Further investigations are still required to fully quantify the impact of SLC 
on weak lensing modelling and weak lensing cosmology.  
(1) Most existing works do not focus on SLC in the context of lensing tomography with realistic width of source redshift.  
These works either study SLC for flux-limited source galaxies with
wide redshift distribution
(e.g. \citealt{Bernardeau98,Hamana01,Hamana02,Schneider02,Romero07}) 
or for fixed source redshift (e.g. \citealt{Schmidt09,Valageas14}).  
The impact of SLC for more realistic situation should fall somewhere between the two extreme cases.  
The SLC caused by the intrinsic source clustering increases with the source redshift width, 
since the cross correlation between the source and lens increases with the source redshift width.  
So this kind of SLC effect can be reduced by lensing tomography with finite redshift width.  
Indeed, in the limit of infinitesimal source redshift width, the SLC effect caused by the intrinsic source clustering 
can be completely eliminated by the standard estimator \citep{Schmidt09}.  
However, photo-z errors and finite galaxy number density prohibit the real redshift width to be smaller than $\sim 0.2$.  
A question to address is then whether in this more realistic situation
lensing tomography can suppress the SLC effect to a level negligible.
(2) Most existing works calculate SLC approximately.  
Approximations made include Taylor expanding the cosmic shear estimator 
and approximations in evaluating high order density correlations.  
An exception is \cite{Romero07}, which analyzed SLC through N-body simulations and hence avoided these uncertainties.  
But \cite{Romero07} only considered the pixel-based estimator, 
only on 1-point statistics (PDF, variance, etc.), and only for flux-limited galaxies.
To make the investigation complete, another estimator (the standard
estimator) should also be studied. 
(3) Given clear evidences that SLC can not be neglected for precision lensing cosmology, 
an immediate task is to calibrate this effect. 
For this purpose we shall quantify its dependence on the galaxy bias or
the magnification prefactor $g$ in this paper. We find
useful scaling relations to correct for the SLC effect. 

We will then study the SLC effect through N-body simulation, for photo-z bins of width $\Delta z^P=0.2$.  
We will consider both sources of SLC, the intrinsic source clustering and the cosmic magnification.
Furthermore, we will consider two popular estimators of shear correlation function/power spectrum.  
(1) The standard estimator.  Cosmic shear is decomposed into the
tangential and cross component for each galaxy pair. Then it sums over the two products of shear components over all pairs of fixed angular separation 
and normalizes by the number of pairs to directly obtain the two point statistics of cosmic shear. 
This standard estimator is widely used in weak lensing surveys.  
(2) The pixel-based estimator.  
It constructs shear maps by averaging over all shears measured in each pixel.  
One then performs the E-B decomposition and measures the 2-point
statistics from these maps. This estimator is convenient for
map-making, one point statistics such as peak statistics and PDF, or
high-order statistics.    The two estimators differ in the order of normalization over the
galaxy number density and averaging over galaxy pairs.   
However, none of
them can satisfactorily handle both sources of SLC. Fortunately we
find that there exist excellent scaling relations between SLC and observables, which
allow for self-calibrating the SLC effect. 

Our paper is organized as follows.  
The SLC effect is described in \S \ref{sec:SLC} for both the standard estimator and the pixel-based estimator.
The way we estimate the SLC effect through simulation is also described in this section,
while the simulation details are attached in the Appendix.
The results are presented in \S \ref{sec:result1} \& \ref{sec:result2} for the standard estimator 
and the pixel-based estimator, respectively.  
We discuss and conclude in \S \ref{sec:conclusion}. For busy readers,
please refer to Table \ref{Table:outline} for figures of major results and
corresponding text. 

\begin{table}[htb]
\caption{The source-lens clustering for different estimators and for
  difference sources. Busy readers can directly refer to corresponding
figures for major results and corresponding sections for explanation
and discussion. Overall we find $\mathcal{O}(1)$-$\mathcal{O}(10)\%$
impact at typical scale $\ell \sim 10^3$ and typical source redshift
$z_s\sim 1$. }
\begin{center}
\begin{tabular}{|r||c||c|}\hline
Source-lens     & The standard &  The pixel-based\\
 clustering (SLC)     & estimator &  estimator\\\hline\hline
Intrinsic   & \S \ref{sec:clustering1} & \S \ref{sec:clustering2}\\
clustering & Fig. 3, 4, 5   & Fig. 8a, 9, 10\\\hline\hline
Cosmic & \S \ref{sec:magnification1} & \S \ref{sec:magnification2}\\
magnification & Fig. 3, 4, 5, 7 & Fig. 8b, 9 \\ \hline
\end{tabular}
\end{center}
\label{Table:outline}
\end{table}


\section{The SLC effect}
\label{sec:SLC}

\bfi{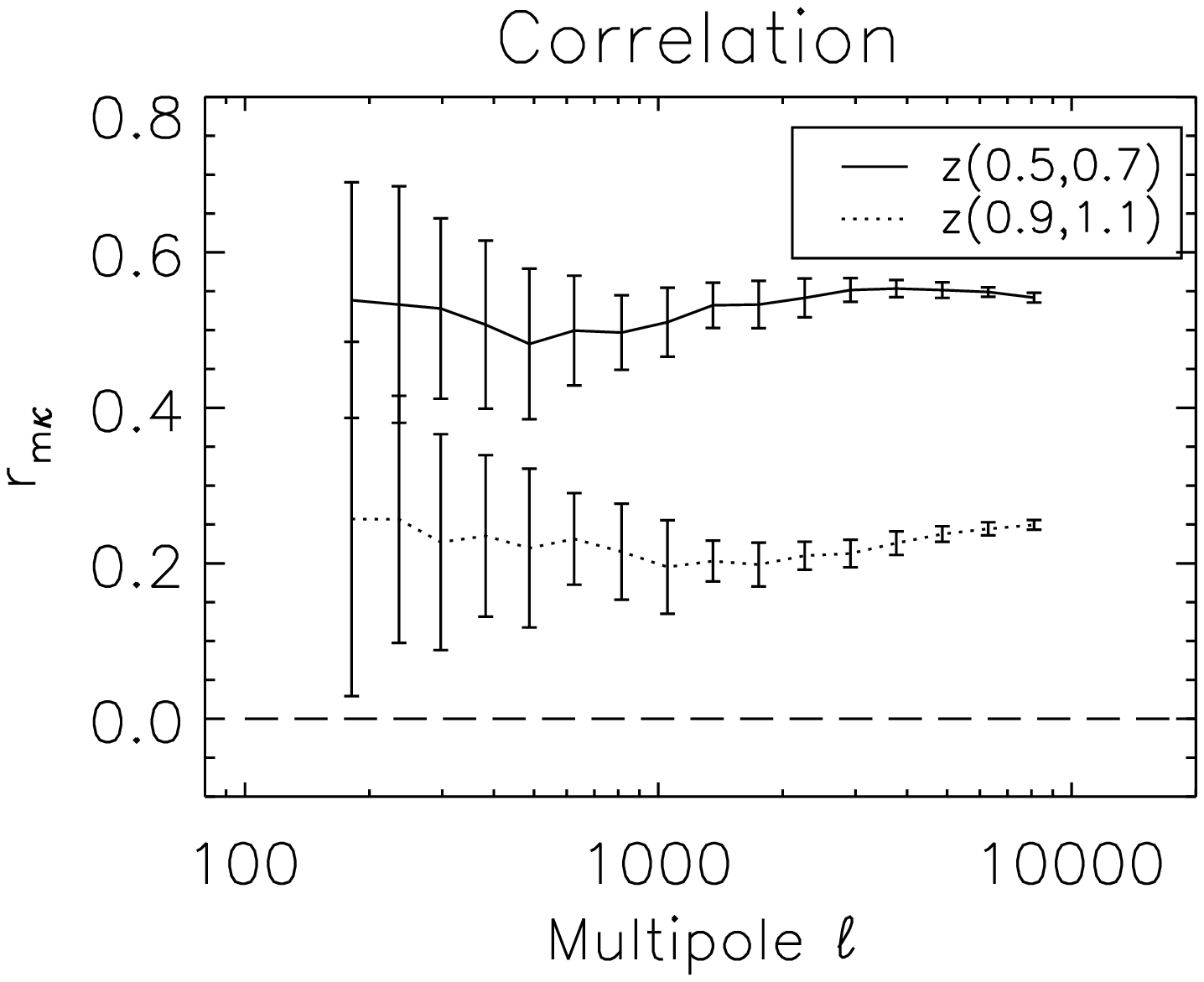}
\caption{Cross correlation coefficient $r_{m\kappa}$ between the source galaxy intrinsic
  distribution ($b_g=1$) and the lensing signal. The error bars are
the rms dispersion among 20 realizations.    In the limit $\sigma_P
\rightarrow 0$ and $\Delta 
  z^P\rightarrow 0$, $r_{m\kappa}\rightarrow 0$. But for more realistic cases
  shown ($\Delta z^P=0.2$ and 
  $\sigma_P=0.05(1+z)$),  $r_{m\kappa}\sim 0.2$-$0.6$ at $\ell \sim 1000$, meaning
  significant spatial  correlation between the source galaxies and the
  lenses. This implies a significant SLC effect, as verified later in
  the paper.  Increasing $r$ with decreasing redshift implies stronger
  SLC at lower redshift, again verified later in the paper. Notice
  that this is just one source of  SLC. The other source is the cosmic
  magnification, which is always perfectly correlated with the lensing
  signal and lensing tomography does not help handle it. 
\label{fig:rmk}} \efi

Weak lensing cosmology prefers volume weighted lensing measurement.  
However, in reality we can measure cosmic shear $\gamma$ only where there are galaxies.  
Thus the measured shear field is inevitably weighted by some function of $n_g$, 
the number density of observed source galaxies.  
The galaxy distribution fluctuates spatially, $n_g=\bar{n}_g(1+\delta^{\obs})$.  
The limit of no correlation between $\delta^{\obs}$ and the lensing field ($\kappa$ and $\gamma$)
corresponds to randomly sampling the lensing field.  
This is a fair sampling process and does not bias the measured lensing statistics.
{\it Unfortunately, in reality $\delta^{\obs}$ correlates with the lensing field}.  
So the lensing field is not fairly sampled observationally.  
Hence this source-lens clustering (SLC) effect biases the lensing measurement.

\subsection{The origins of SLC}
The observed source galaxy number overdensity $\delta^{\obs}$, to
leading order, has two components,
\be
\delta^{\obs}=\delta_g+g\kappa .  
\ee 
Here, $\delta_g$ is the
intrinsic fluctuation (intrinsic clustering).  Throughout the paper we
adopt a simple bias model for $\delta_g=b_g\delta_m$ where $\delta_m$
is the matter overdensity and $b_g$ is the galaxy bias.
$g\kappa$ is that induced by
the cosmic magnification.  It arises from the thresholds in selecting galaxies, either
by flux (magnification bias) or size (size bias).

Since SLC arises from $\delta^{\rm obs}$-$\kappa$ ($\gamma$) correlation, 
its strength depends on the cross correlation coefficient 
$r_{\rm SL}=C_{\delta^\obs \kappa}/\sqrt{C_{\delta^\obs\delta^\obs}C_{\kappa\kappa}}$ between the two.  
Here the subscript "SL'' denotes ``source-lens''.  
$C_{\delta^\obs \kappa}$, $C_{\delta^\obs\delta^\obs}$ and $C_{\kappa\kappa}$ are the cross- and auto-power spectrum
between the observed galaxy distribution and lensing convergence.
For constant galaxy bias $b_g$ and magnification prefactor $g$,
\begin{equation}
\label{eqn:rmk}
r_{\rm SL}(\ell)=\left\{
\begin{array}{ll}
\frac{C_{m\kappa}(\ell)}{\sqrt{C_{mm}(\ell)C_{\kappa\kappa}(\ell)}}\equiv r_{m\kappa}\ ,
& \delta^\obs=b_g\delta_m\ , \\
1\ , & \delta^\obs=g\kappa\ .
\end{array}
\right.
\end{equation}
Here $r_{m\kappa}$ is the cross correlation coefficient between the
matter distribution and lensing convergence of the source redshift bin.

In Fig. \ref{fig:rmk}, we plot $r_{m\kappa}$ measured through
simulation, for two typical photo-z bins, $z^P\in (0.5, 0.7)$ and
$(0.9, 1.1)$.  
We adopt a source distribution 
\be 
n^P(z^P)\dif z^P=\frac{1}{2}\frac{z^P}{z_0^3}\exp(-\frac{z^P}{z_0})\dif z^P 
\ee 
with $z_0=0.45$, typical  for LSST-like surveys.  
We assume the photo-z scatter is perfectly known to be
in a Gaussian form with photo-z error $\sigma_P=0.05(1+z)$.
In reality lensing survey also suffers from catastrophic errors.
The safety line for unbiased dark energy parameters constraints was studied in \cite{Hearin10}.
Also several facts prevent us from accurately knowing the scatter function $P(z^P|z)$ \citep{Cunha12a,Cunha12b}.
However, our assumption captures the main feature of the photo-z error except the catastrophic error.
The real galaxy distribution contributing to photo-z bin $(0.5,0.7)$ and $(0.9,1.1)$ is shown in Fig. \ref{fig:zdist}.  
Photo-z errors broaden the redshift distribution and cause substantial overlap between the source and lens distribution.  
This results in a significant $r_{m\kappa}\sim 0.2$-$0.6$ at $100\la
\ell\la 10^4$.   The correlation strength is stronger towards lower redshift, because of the stronger redshift dependence of lensing kernel.
This result directly indicates larger SLC effect due to the intrinsic source
clustering towards lower redshift \footnote{We should point out here that the correlation coefficient does not
completely determine the strength of SLC. For example, it  is
invariant to a deterministic galaxy bias.  But sources with stronger clustering (larger $b_g$) has larger SLC effect.}.
The correlation coefficient only shows weak dependence on the scale, 
indicating that the SLC effect can be important at all scales.

\bfi{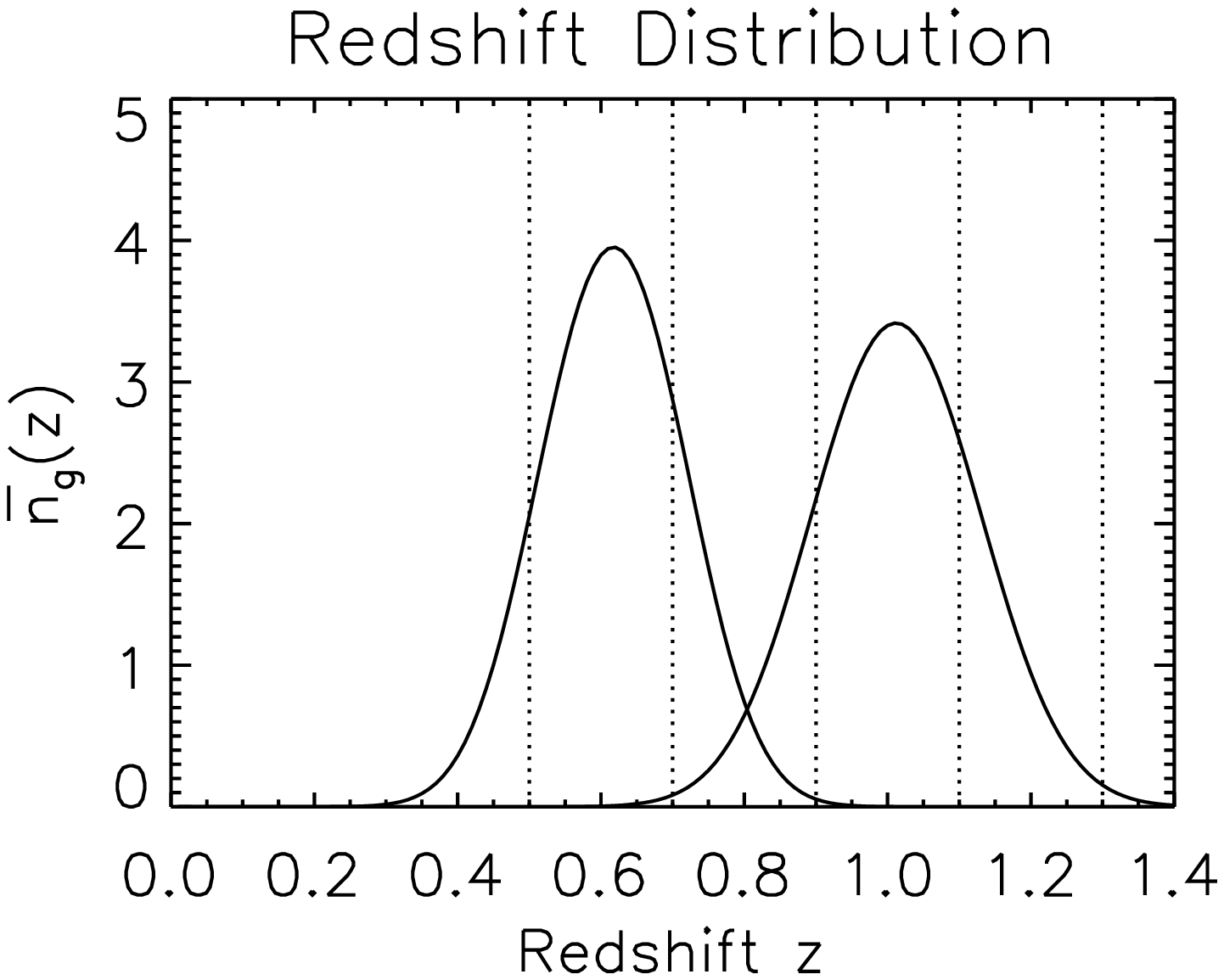}
\caption{The real source galaxy distribution $\bar{n}_g(z)$ of photo-z bins
$(0.5,0.7)$ and $(0.9,1.1)$ are presented.  $\bar{n}_g(z)$ is normalized to
$\int \bar{n}_g(z) \dif z=1$.  Photo-z error broadens the redshift
distribution significantly, and hence increases the SLC effect. 
\label{fig:zdist}} \efi

Narrowing the photo-z bin reduces $r_{m\kappa}$.  
But this reduction is very mild, because the redshift width is $\sim 2\sigma_P\sim 0.2$ even in the limit $\Delta z^P=0$ for $z_s \sim 1$.
Hence even with the aid of photo-z information, SLC induced by the
intrinsic source clustering can not be significantly suppressed.  
We then expect non-negligible SLC even in the context of lensing tomography.

The exact impact of SLC depends on the estimator of these lensing statistics.  
In this work we investigate two widely
used estimators for the 2-point correlation function and power
spectrum.

\subsection{The standard estimator} 
\label{subsec:se}
The standard estimator \citep{Munshi08,Schmidt09,CFHTLenS} averages shear products on
galaxy pairs in the lensing survey.  For each galaxy cosmic shear is
decomposed into the tangential shear $\gamma_\rmt$ and the cross
component $\gamma_\rmx$ relative to the separation vector to the other
galaxy in the pair.  Then one calculates the following shear-shear
correlation functions: 
\be
\label{eqn:estimator1} 
\xi_\pm(\theta)= \frac{\sum_{\alpha\beta}
[\gamma_{\rmt,\alpha}\gamma_{\rmt,\beta}
\pm\gamma_{\rmx,\alpha}\gamma_{\rmx,\beta}]
w_{\alpha}w_{\beta} \Delta_{\alpha\beta}}
{\sum_{\alpha\beta} w_{\alpha} w_{\beta}\Delta_{\alpha\beta}}\ , 
\ee 
where galaxy pairs labeled $\alpha$, $\beta$ are separated by angular distance $\vartheta=|\vartheta_\alpha-\vartheta_\beta|$.  
$\Delta_{\alpha\beta}=1$ counts for all the pairs with $\vartheta$
falling in the angular bin centered at $\theta$, 
otherwise $\Delta_{\alpha\beta}=0$.  
$w_{\alpha}$ is the weighting for each galaxy depending on some
selection factors like the image quality, etc..   For brevity, we only
discuss the simplified case $w_\alpha=1$.  Ideally (e.g. in the absence of SLC), these correlation functions are
connected to the theory through the relationship  
\be
\label{eqn:e2theory} 
\xi_{+/-}(\theta)=\frac{1}{2\pi}\int_0^\infty\dif \ell\ell J_{0/4}(\ell\theta)C_{\kappa\kappa}(\ell)\ ,
\ee 
in which $J_{0/4}$ is the 0th- and 4th-order Bessel function of the first kind.
However, as shown above, the SLC effect is implicitly included through the
summation over all galaxy pairs.   

E-B decomposition is very useful in weak lensing analysis.  
B-modes are strong evidence for systematic errors, 
such as the point spread function \citep{Hoekstra04}, shear calibration errors, 
galaxy intrinsic alignment, finite volume effect \citep{Kilbinger06},
and the SLC effect \citep{Schneider02}.  
From theoretical aspect, the E/B-mode power spectra are related to the correlation functions as 
\be
C_{E/B}(\ell)=\pi\int_0^\infty\dif\theta\theta[J_0(\ell\theta)\xi_+(\theta)\pm J_4(\ell\theta)\xi_-(\theta)]\ .  
\ee

In simulation implementation, the calculation of E/B-mode power spectrum can be
simplified, without calculating $\xi_{\pm}$. 
First one measures the shear correlation functions
\be
\begin{split}
\label{eqn:standard}
&\hat\xi_{ij}(\vec{\theta})= \\
&\frac{\langle\int\dif z\bar{n}_g(z)(1+\delta^\obs(z))\int\dif z'\bar{n}_g(z')(1+\delta^\obs(z'))\gamma_i(z)\gamma_j(z')\rangle}
{1+\xi_{\delta^\obs}(\theta)}\ ,
\end{split}
\ee
in which the denominator $(1+\xi_{\delta^\obs}(\theta))=\langle\int\dif z\bar{n}_g(z)(1+\delta^\obs(z))\int\dif z'\bar{n}_g(z')(1+\delta^\obs(z'))\rangle_{\theta=|\vec\theta|}$ is the correlation function of the observed galaxy distribution.
It averages over direction of $\vec\theta$.  But the numerator does not. 
$i,j=1,2$ denote for the shear components defined related to axes. 
The prime denotes for another line-of-sight with separation $\vec\theta$.
The integral (summation) is over all galaxies in the bin.  This allows us to do a 2D Fourier transform to
obtain the corresponding 2D power spectrum $C_{ij}(\vec{\ell})={\rm FFT}(\hat\xi_{ij}(\vec{\theta}))$. We
then average over the direction of $\vec\ell$ to obtain the E/B-mode power
spectrum, 
\be
\label{eqn:P_EB}
\begin{aligned}
C_E(\ell)&=\langle C_{11}(\vec\ell)\cos^{2}2\varphi_{\vec\ell}\rangle
+\langle C_{22}(\vec\ell)\sin^{2}2\varphi_{\vec\ell}\rangle\\ 
&+\langle (C_{12}(\vec\ell)+C_{21}(\vec\ell))\cos2\varphi_{\vec\ell}\sin2\varphi_{\vec\ell}\rangle\ ,\\
C_B(\ell)&=\langle C_{11}(\vec\ell)\sin^{2}2\varphi_{\vec\ell}\rangle
+\langle C_{22}(\vec\ell)\cos^{2}2\varphi_{\vec\ell}\rangle\\ 
&-\langle (C_{12}(\vec\ell)+C_{21}(\vec\ell))\cos2\varphi_{\vec\ell}\sin2\varphi_{\vec\ell}\rangle\ .
\end{aligned}
\ee 
Here $\varphi_{\vec\ell}$ is the angle between the vector $\vec\ell$
and the x-axis in 2D Fourier space.

\subsection{The pixel-based estimator}
\label{subsec:pe}
For the pixel-based estimator \citep{Schmidt09}, one directly obtains an averaged cosmic
shear $\gamma$ from all galaxies in each pixel,
\be
\label{eqn:pixelbased}
\hat\gamma_{i}=\frac{\int\bar{n}_g(1+\delta^\obs)\gamma_{i}\dif z}{\int\bar{n}_g(1+\delta^\obs)\dif z}\ .
\ee
 Then we can construct
lensing convergence map through 
\be
\label{eqn:kappa_E}
\kappa_E(\vec\ell)=\hat\gamma_1(\vec\ell)\cos2\varphi_{\vec\ell} +
\hat\gamma_2(\vec\ell)\sin2\varphi_{\vec\ell}\ .  
\ee 
The parity asymmetric B-mode convergence $\kappa_B$ could
be constructed as 
\be
\label{eqn:kappa_B}
\kappa_B(\vec\ell)=\hat\gamma_1(\vec\ell)\sin2\varphi_{\vec\ell}
-\hat\gamma_2(\vec\ell)\cos2\varphi_{\vec\ell}\ .
\ee
Then we could obtain E/B-mode power spectrum
$C_{E}(\ell)=\langle\kappa_{E}(\vec\ell)\kappa_{E}^*(\vec\ell)\rangle$ and
$C_{B}(\ell)=\langle\kappa_{B}(\vec\ell)\kappa_{B}^*(\vec\ell)\rangle$.

\subsection{Different estimator, different SLC} 

The impact of SLC effect on these two estimators could be figured out in extreme cases. 

For the standard estimator Eq. \ref{eqn:standard}, 
if we have accurate redshift measurement and infinitesimal redshift bin size,
the correlation between intrinsic source clustering and lens will be eliminated,
$\langle\delta_g\delta_g'\gamma_i\gamma_j'\rangle\rightarrow
\langle\delta_g\delta_g'\rangle\langle\gamma_i\gamma_j'\rangle$.  
Thus in this case the normalization will perfectly cancel the SLC induced by the intrinsic source clustering,
$\hat\xi_{ij}\rightarrow\xi_{ij}\equiv\langle\gamma_i\gamma_j'\rangle$.  
However, in the context of lensing tomography with photo-z measurement, 
there exists significant correlation between the source and lens (Fig. \ref{fig:rmk}).
The investigation in this paper will quantitively give us a conclusion about how large the SLC effect is 
and its dependence on redshift, scale and galaxy bias. 
For the cosmic magnification, $\delta^\obs=g\kappa$ always perfectly correlates with the lensing signal.  
Thus inevitably the standard estimator will suffer from the cosmic magnification.
Adopting analytical analysis, the amplitude is expected to be proportional to 
$g\langle\kappa\gamma_i\gamma_j'\rangle/\langle\gamma_i\gamma_j'\rangle$.

For the pixel-based estimator Eq. \ref{eqn:pixelbased}, 
in the limit of constant $g\kappa$ across the photo-z bin, 
$\int(1+g\kappa)\gamma_i\dif z\rightarrow(1+g\kappa)\int\gamma_i\dif z$.
Thus the normalization for each pixel will cancel the cosmic magnification induced SLC,
leading to $\hat\gamma_i\rightarrow\gamma_i$ and $\hat\xi_{ij}\rightarrow\xi_{ij}$.
The following investigation will tell us the SLC induced by the cosmic magnification is indeed negligible.
However, the intrinsic source clustering is not a smooth function of redshift.
Even worse, for low redshift the amplitude is not weak enough to make analytical analysis.
Simulation is of great help to investigate this problem.

\bfi{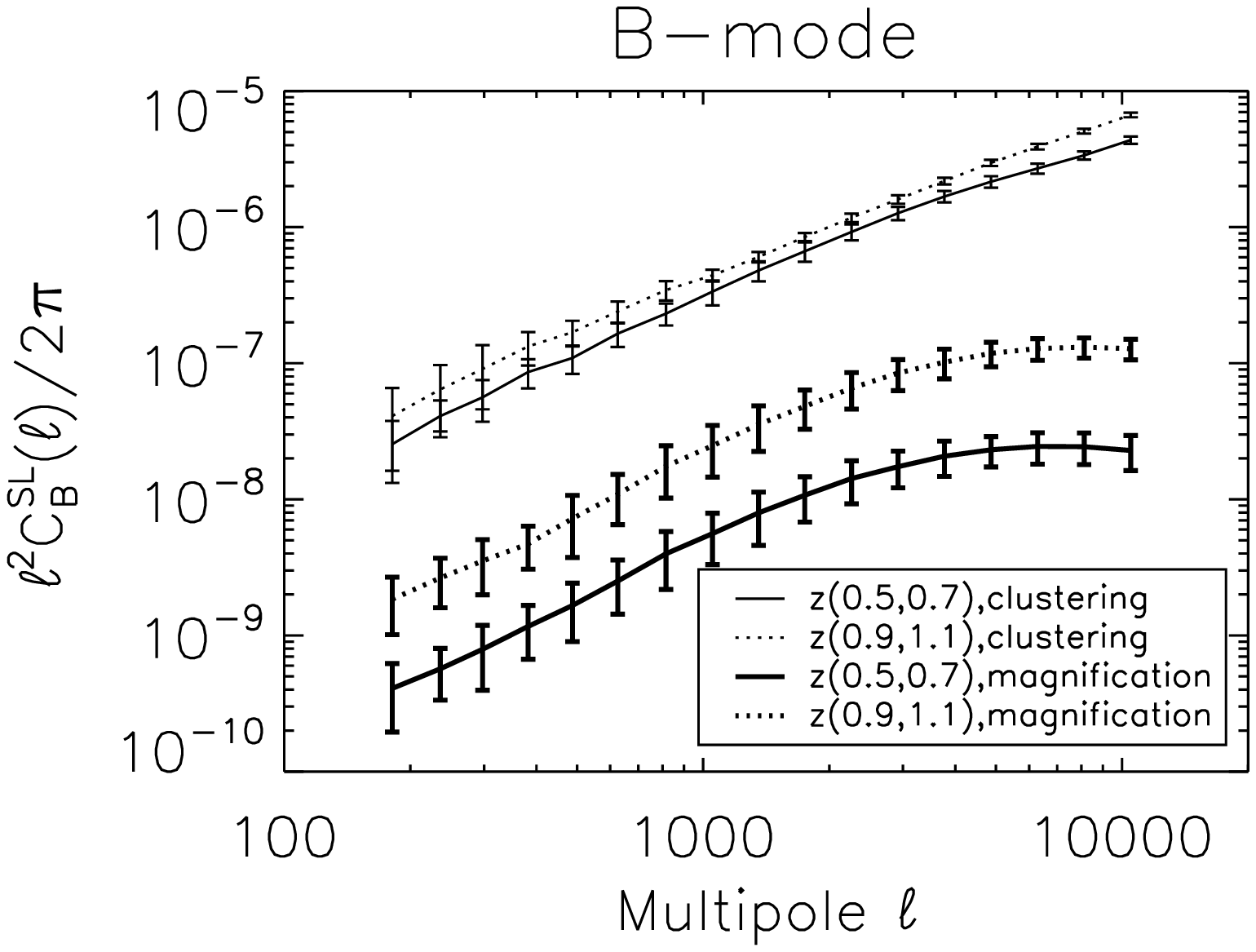}
\caption{SLC induced B-mode power spectrum in the standard estimator,
  for the case of intrinsic 
  source clustering (thin lines) and cosmic magnification (thick
  lines). $b_g=1$ and $g=1$ are adopted. Solid and dotted lines are
  for photo-z bins $(0.5,0.7)$ and $(0.9,1.1)$, respectively. 
The error bars are the rms dispersion among 20 realizations.
The B-mode is robustly identified in all cases.
\label{fig:gdbmode}} 
\efi

\subsection{Quantification}
Through N-body simulation, we can construct 2D maps of source and lens
at various redshifts to robustly quantify the SLC effect. We adopt the
Born approximation for lensing map construction. The SLC
effect persists under the Born approximation, since at least it keeps
the leading order clustering between sources and lenses. Hence it is
sufficient to adopt the Born approximation to study the SLC
effect. This allows us to construct lensing maps by stacking randomly
rotated and shifted simulation boxes. For details of map
construction, we refer the readers to the Appendix. With these
simulated maps we are able to calculate the SLC straightforwardly. 

We consider two typical photo-z bins $(0.5,0,7)$ and $(0.9,1.1)$. This
allows us to
investigate the dependence of SLC on source redshift. We investigate
two measures of SLC. (1) One is the B-mode power spectrum $C_B^{\rm
  SL}(\ell)$. Without SLC, the simulated cosmic shear has vanishing
B-mode. So $C_B^{\rm SL}(\ell)$ provides a useful measure of SLC. (2)
The other is the ratio of the E-mode power spectra in the presence
and in the absence of SLC effect, 
\be
\label{eqn:psratio}
R(\ell)\equiv\frac{C_{E}^{\rm SL}(\ell)}{C_{E}(\ell)}\ .
\ee
We measure it for the two auto-power
spectra and one cross-power spectrum between the two redshift bins. 

We are also interested in the dependences of SLC on the galaxy bias
$b_g$ and magnification prefactor $g$. Understanding these
dependences helps calibrate the SLC effect. 
We choose $b_g=0.5,1,1.5,2$ and $g=-1,1,1.5,2$ to study these dependences.
We just substitute different $b_g$ and $g$ values in the construction of $\delta^\obs(\hat{n},z)$. 
To avoid unrealistic $\delta^\obs(\hat{n},z)<-1$, which could happen if $b_g>1$ or $g\gg1$, 
we simply set $\delta^\obs(\hat{n},z)=-1$ where it happens. 
\begin{figure*}
\epsfxsize=16cm
\epsffile{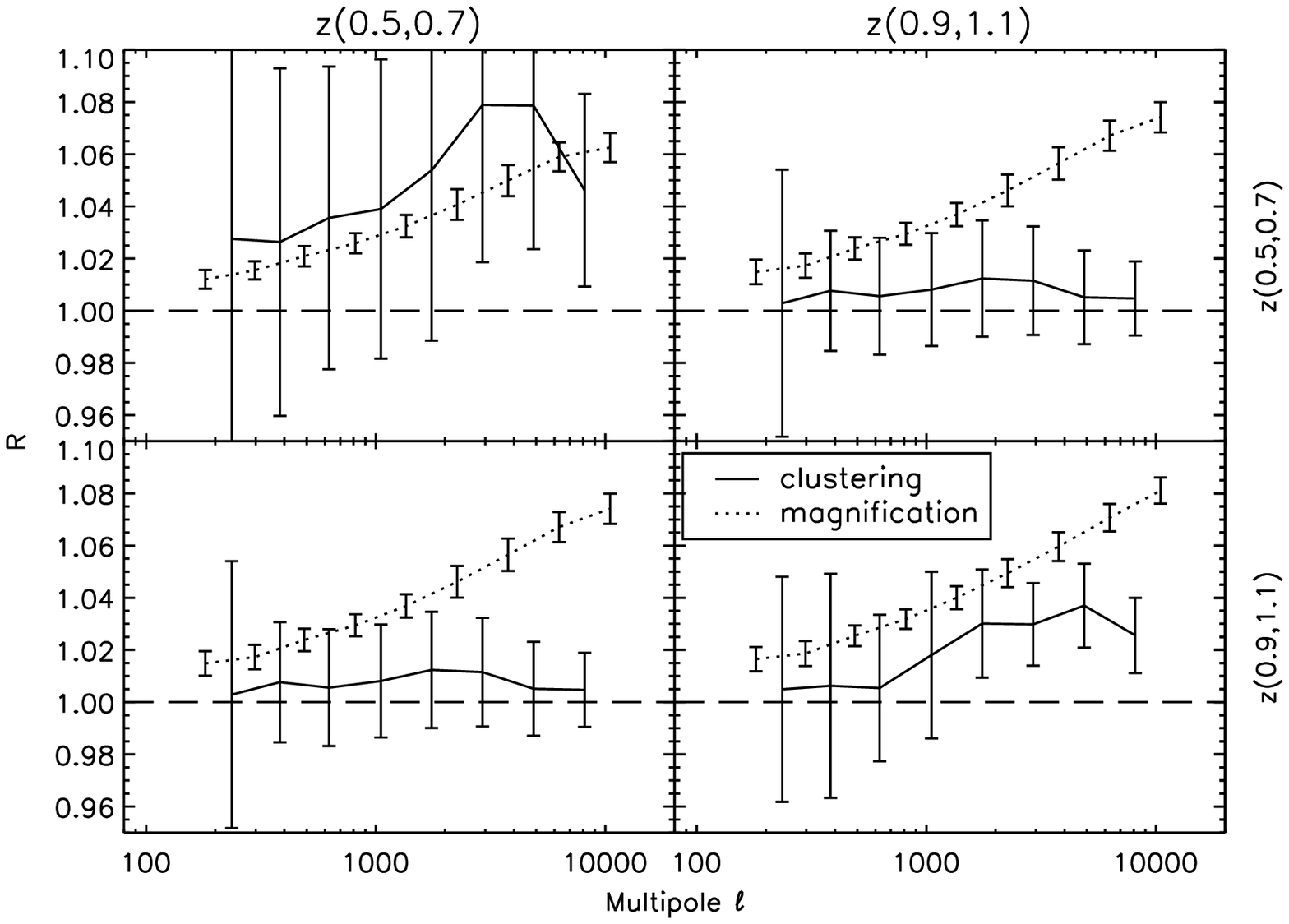}
\caption{The impact of SLC on the lensing
  auto- and cross-power spectrum measured by the standard
  estimator. We plot the ratio of the power  spectra $R=C_{E}^{\rm
    SL}(\ell)/C_{E}(\ell)$, for $b_g=1$ and $g=1$, and for photo-z bins
  $(0.5,0.7)$ and $(0.9,1.1)$.  $512^3$ uniform grid is adopted in the calculation.
  The error bars are estimated from 20
  realizations.  SLC can bias the lensing E-mode power
  spectrum measurement by  $1\%$-$8\%$ and the exact value depends on
  the galaxy bias ($b_g$) and the magnification pre-factor ($g$).
\label{fig:gdcps}}
\end{figure*}

\section{SLC with the Standard Estimator}
\label{sec:result1}

This section presents our results for the impact of  SLC on lensing power spectrum
measured with the standard estimator.

\subsection{Intrinsic Source Clustering}
\label{sec:clustering1}

We take $\delta^\obs=b_g\delta_m$ in Eq. \ref{eqn:standard} to quantify
SLC effect due to the intrinsic source clustering. We adopt $b_g=1$ first.

\subsubsection{B-mode power spectrum}
The B-mode power spectrum is presented in Fig. \ref{fig:gdbmode},
which is about 3 orders of magnitude lower than the E-mode.  Despite its small
amplitude, the measurement of B-mode is robust, since its amplitude is 
5 orders of magnitude larger than the case neglecting SLC effect, i.e. setting $\delta^\obs=0$
(which is not presented here).    The B-mode amplitude is larger for
higher photo-z bin $(0.9,1.1)$.  But this does not mean that the SLC effect is larger.
Direct comparison of B-mode at the two redshifts is less meaningful, 
since the E-mode signal is also larger for higher redshift.

\begin{figure*}
\epsfxsize=8cm
\epsffile{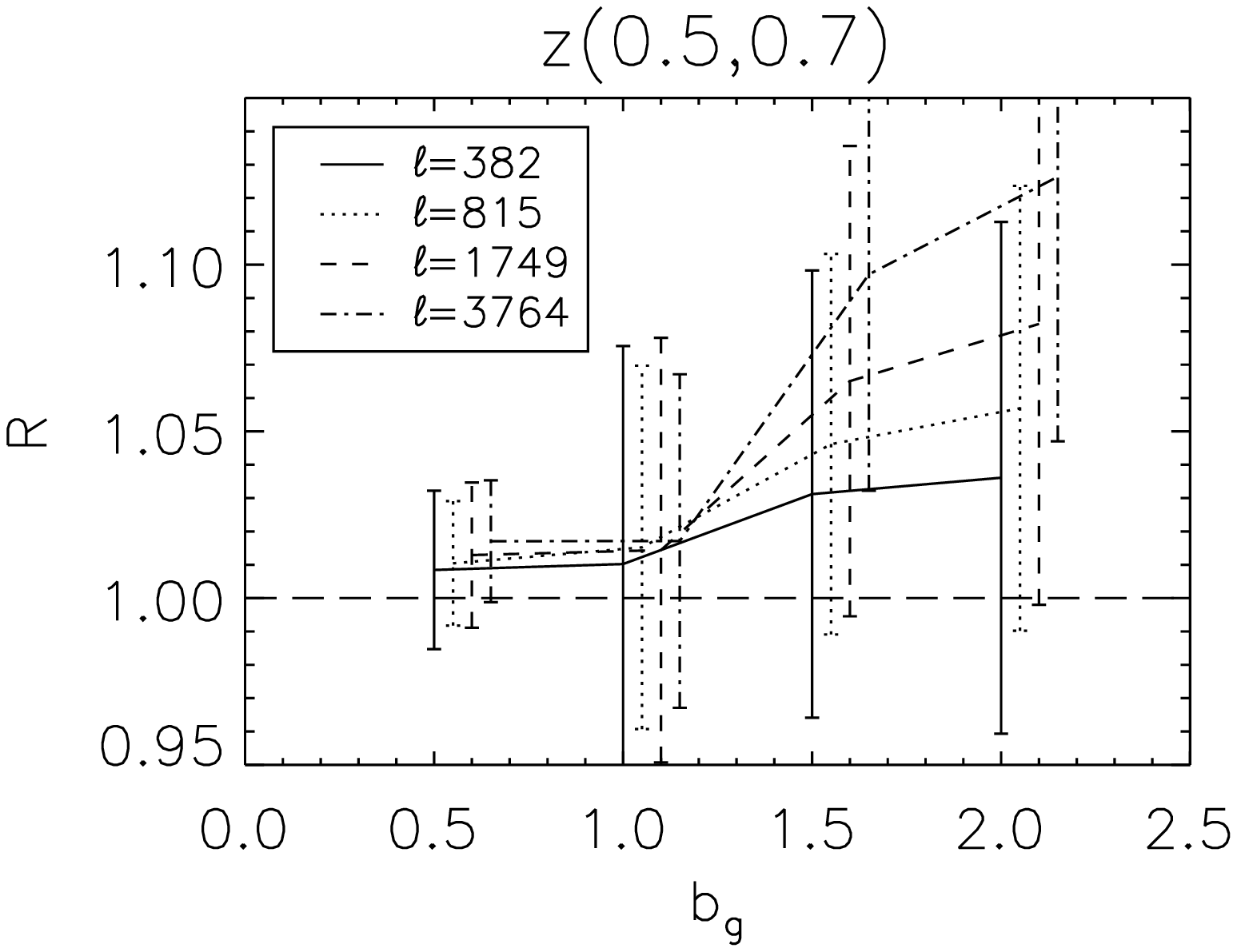}
\epsfxsize=8cm
\epsffile{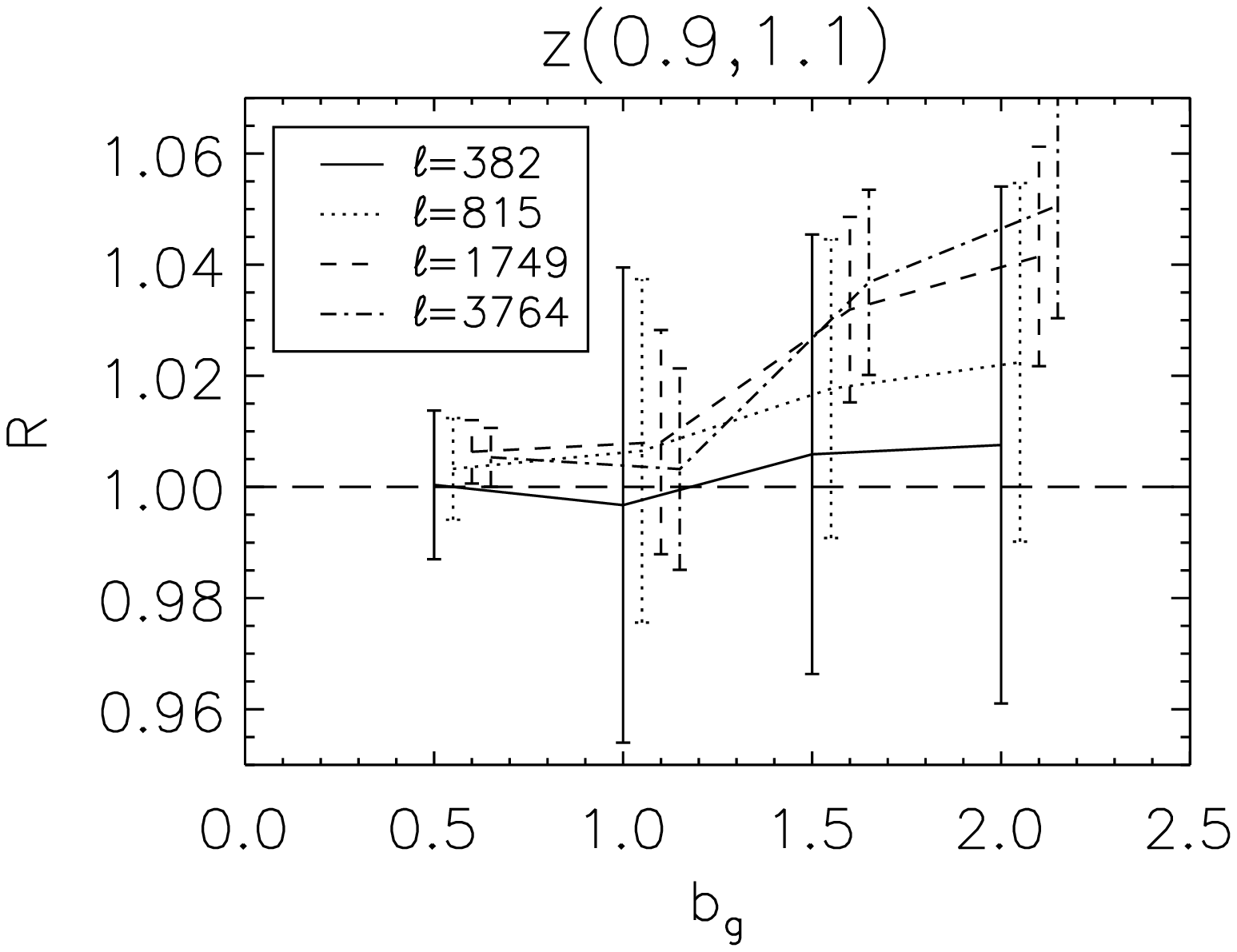}
\caption{The dependence of $R=C_{E}^{\rm SL}(\ell)/C_E(\ell)$ on
  $b_g$ for standard estimator. The data points have  $b_g=0.5,1,1.5,2$ respectively. 
For clarity, the results for different $\ell$'s are presented with
arbitrary horizontal shifts.
The dependence on galaxy bias is nonlinear.  
\label{fig:gdbb}}
\end{figure*}

\subsubsection{E-mode power spectrum}
The impact of  SLC on the E-mode auto- and cross-power spectrum,
quantified by the ratio $R$  (Eq. \ref{eqn:psratio}),  is
shown in Fig. \ref{fig:gdcps}. Although the measurement on $R$ has
large statistical uncertainty, we find $R>1$. For $b_g=1$, SLC
amplifies the lensing E-mode power spectrum by $1$-$8\%$ at $100\la
\ell \la 10^4$ for the photo-z bin $(0.5,0.7)$. The impact becomes
weaker at the higher photo-z bin $(0.9,1.1)$. But its amplitude is
still of the order $1\%$ and hence non-negligible for precision weak
lensing cosmology. 

This overestimation of the E-mode lensing power spectrum by the
standard estimator is caused by the positive cross correlation coefficient between the
lensing signal and the galaxy intrinsic clustering (See Fig. \ref{fig:rmk} and Eq. \ref{eqn:rmk}).
Although in \cite{Valageas14} negative correlation functions are found in some high-order correlation terms at large scales,
it is not in contradiction with our result of positive cross correlation coefficient on all scales.
One difference between the two is that our approach includes all the high-order corrections.
The other difference is that we describe the SLC in terms of power spectrum ratio instead of correlation function ratio.
In general cases, correlation functions become negative at large scales due to the normalization.
We prefer to use power spectrum since the errors in modes are independent at large scales or under linear evolution.
However the description and explanation using correlation functions is more straightforward and in principle they are consistent.
From
Eq. \ref{eqn:standard}, 
\be 
\label{eqn:Rminus1}
\hat{\xi}_{ij}-\xi_{ij}=\frac{\langle
  (\delta_g+\delta_g^{'})\gamma_i\gamma^{'}_j\rangle+\langle\delta_g\gamma_j^{'}\rangle\langle\delta_g^{'}\gamma_i\rangle+\langle\delta_g\gamma_i\delta_g^{'}\gamma_j^{'}\rangle_c}{1+\langle
  \delta_g\delta_g^{'}\rangle}\ .
\ee
Since the intrinsic galaxy distribution  is positively correlated with the lensing
signal, the standard estimator overestimates the lensing power
spectrum and we expect $R-1>0$.

In contrast, the impact of SLC on the lensing cross-spectrum is much weaker and the
measured $R-1$ is consistent with zero, within the error bars. For
this case, we have
\be 
\hat{\xi}_{ij}-\xi_{ij}=\langle
  (\delta_g+\delta_g^{'})\gamma_i\gamma^{'}_j\rangle\ .
\ee
Hence we expect a weaker SLC effect. However, the above result shows
that the SLC effect on cross-power spectrum is non-zero, although we
lack the statistical accuracy to measure it. 

We find that the standard estimator results in large cosmic variance in the
power spectrum measurement (Fig. \ref{fig:gdcps}).  For example,
statistical fluctuation in $R$ is  $\sim 5\%$ for the photo-z bin
$(0.5,0.7)$ over the angular scale $100\la \ell \la 10^4$. In
particular, the cosmic variance does not decrease toward smaller
angular scales, as we would expect. This is in sharp contrast with
statistical fluctuation in the true E-mode lensing power spectrum or
the angular galaxy power spectrum (Fig. \ref{fig:cmmckk}). 

We believe that this large cosmic variance is the result of modulation
of the denominator  ($1+\xi_{\delta_g}$) in the standard estimator
(Eq. \ref{eqn:standard}). Since the correlation between the lensing signal and the galaxy
field is not strong, statistical fluctuation in the galaxy field
($\delta\xi_{\delta_g}$) is only partly
eliminated in the standard estimator (Eq. \ref{eqn:standard}). So it
contributes to fluctuation in the measured lensing correlation
function/power spectrum.  Since the galaxy clustering is much stronger
than the lensing signal ($C_{mm}/C_{\kappa\kappa}\sim 10^3$, Fig. \ref{fig:cmmckk}),
statistical fluctuation in the denominator can cause  much
larger (fractional) fluctuation in the measured lensing correlation function/power
spectrum. This also explains the reduced error bars towards higher redshift, 
for the increasing magnitude of lensing signal and decreasing magnitude of $\xi_{\delta_g}$.  

\subsubsection{Dependence on $b_g$}

We neglect the evolution of galaxy bias across the redshift bin we consider.  
The results for $b_g=0.5, 1, 1.5, 2.0$ we arbitrarily choose are shown in
Fig. \ref{fig:gdbb}. The dependence on $b_g$ is clearly nonlinear. This
more complicated behavior is already implied by
Eq. \ref{eqn:Rminus1}.  From it, we have 
\be
R-1=\frac{b_gf_1(\ell)+b_g^2f_2(\ell)}{1+b_g^2f_3(\ell)}\simeq b_gf_1+b_g^2f_2-b_g^3f_1f_3+\cdots\ ,
\ee
in which $f_1,f_2,f_3$ are some functions of the underlying matter density $\delta_m$ 
and the lensing signal $\gamma$.
Since the density fluctuation is of the order unity
(Fig. \ref{fig:cmmckk}), $f_2$ is comparable to $f_1$, leading to
deviation from the linear dependence. 

In principle we can fit the above results to obtain a fitting formula
for the $R$-$b_g$ relation. Such relation is useful to calibrate SLC
in the standard estimator. $b_g$ (up to a normalization) can be
directly measured from the lensing survey. So by splitting source
galaxies into flux bins (with different $b_g$), we can measure
$C_E^{\rm SL}$-$b_g$ relation and extrapolate to obtain $C_E$ in the
limit of $b_g=0$. Unfortunately our simulation does not have
sufficient independent realizations to beat down statistical
fluctuations in $R$ (Fig. \ref{fig:gdbb}), so we postpone more
detailed investigation of $R$-$b_g$ relation into future works. 

More careful analysis should be carried out against mock galaxy
catalogue with galaxies of various types.  Nevertheless, results here
have robustly shown that SLC is non-negligible in the lensing
correlation function/power spectrum measured through the standard
estimator.

\subsection{Cosmic Magnification}
\label{sec:magnification1}

We take $\delta^\obs=g\kappa$ in Eq. \ref{eqn:standard} to quantify
the SLC effect due to the cosmic magnification.
In the analysis,   we neglect the change
of $g$ across the redshift bin we consider. Unless for very bright
galaxies in the exponential tail, this approximation is sufficiently
accurate. Firstly we assume $g=1$. We then try other values of $g$
from $-1$ to $2$. 

\subsubsection{B-mode power spectrum}
The B-mode caused by the cosmic magnification induced SLC  is robustly identified
(Fig. \ref{fig:gdbmode}).   It is 2 orders of magnitude lower than the
one due to the intrinsic source clustering and 5 orders of magnitude
lower than the lensing E-mode.  Hence it is too weak to detect in lensing
surveys. However, this does not mean its impact on the E-mode lensing
power spectrum is negligible. In contrast, its impact on the E-mode
power spectrum is 4 orders of magnitude larger than the amplitude of B-mode itself. As shown in
Fig. \ref{fig:gdcps}, the SLC caused by this
cosmic magnification  affects  the lensing E-mode power spectrum by a
few percent. This is an example  showing the danger of using B-mode as a
diagnostic of lensing systematic errors.    
\bfi{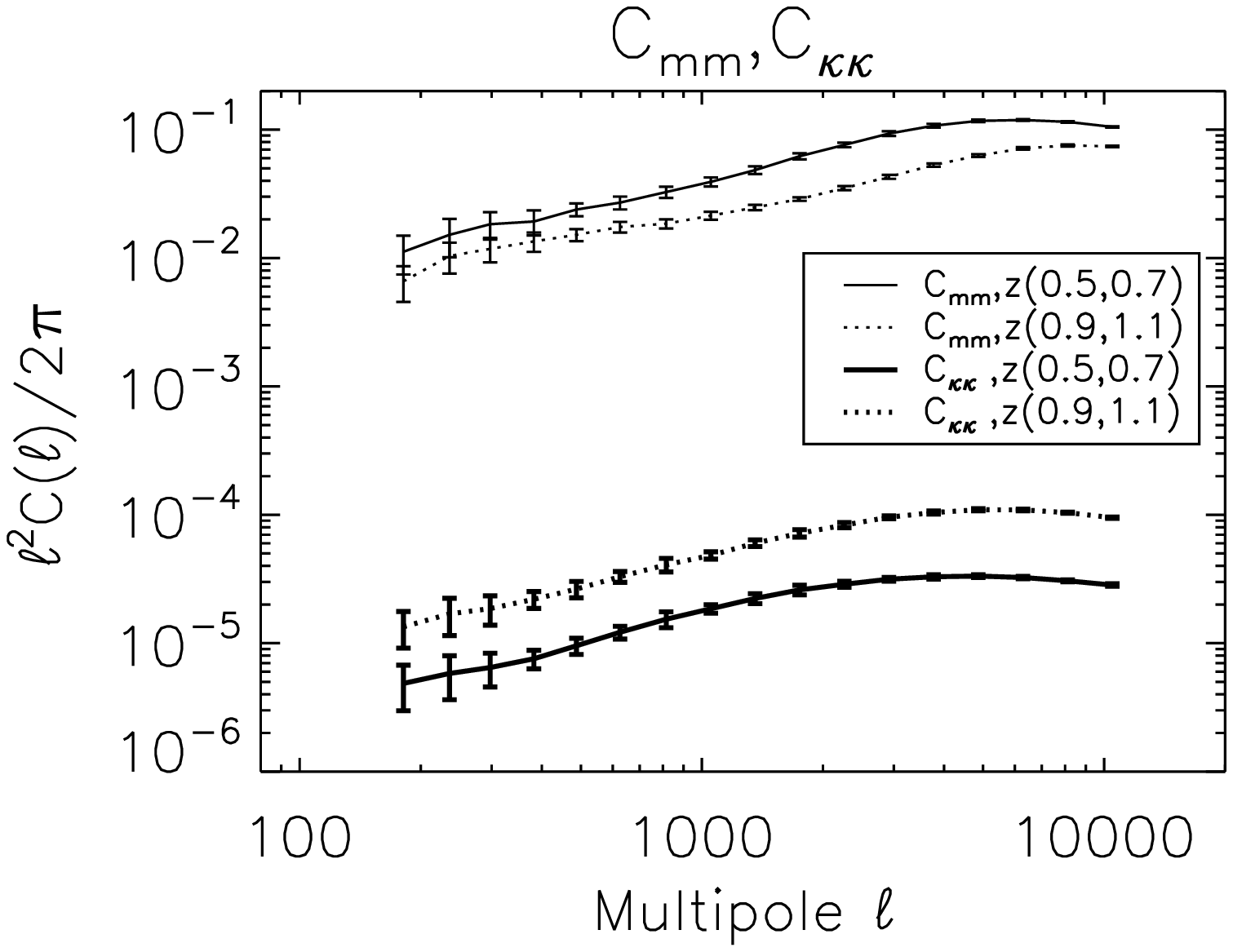}
\caption{The lensing power spectrum $C_{\kappa\kappa}$ and the galaxy angular power
  spectrum with $b_g=1$ ($C_{mm}$) for photo-z bins $(0.5,0.7)$ and $(0.9,1.1)$. Large
  fluctuations in the galaxy distribution enhance statistical
  fluctuations in the lensing power spectrum measured by the standard
  estimator. 
\label{fig:cmmckk}} 
\efi

\subsubsection{E-mode power spectrum}
SLC induced by the cosmic magnification has significant impact on the
lensing E-mode power spectrum (Fig. \ref{fig:gdcps}). For example, at $\ell=1000$
of particular interest in weak lensing cosmology,  it enhances the
lensing power spectrum by 3\% for photo-z bin $(0.5,0.7)$ and 4\% for
photo-z bin $(0.9,1.1)$. The SLC effect on the cross-spectrum falls
in between.  The SLC effect also increases with $\ell$,
from $1\%$ at $\ell\simeq 200$ to $6\%$ at $\ell\simeq 10^4$, for the
photo-z bin $(0.5,0.7)$. These behaviors can be roughly understood as follows. From
Eq. \ref{eqn:standard}, we have (also refer to \cite{Schmidt09})
\ba
\label{eqn:CM1}
\hat{\xi}_{ij}-\xi_{ij}&=& \frac{2g\langle
  \kappa\gamma_i\gamma^{'}_j\rangle+g^2\langle\kappa\gamma_j^{'}\rangle\langle\kappa^{'}\gamma_i\rangle+g^2\langle\kappa\gamma_i\kappa^{'}\gamma_j^{'}\rangle_c}{1+g^2\xi_{\kappa\kappa}}\no
\\
&\simeq &2g\langle
  \kappa\gamma_i\gamma^{'}_j\rangle\ .
\ea
The last expression has adopted the approximation $|\kappa|\ll 1$. The
leading order correction $\langle \kappa\gamma\gamma\rangle$ grows
faster than $\langle \kappa\kappa\rangle$ towards small scale and
high redshift, so $R-1$ increases with both source redshift and
$\ell$ (Fig. \ref{fig:gdcps}). 

In contrast to the case of  the intrinsic source clustering, $R$ among
realizations show weaker fluctuations (Fig. \ref{fig:gdcps}). The
reason is that statistical fluctuations in the denominator and the
numerator of the standard estimator are strongly correlated and hence
are largely cancelled out.  

\bfi{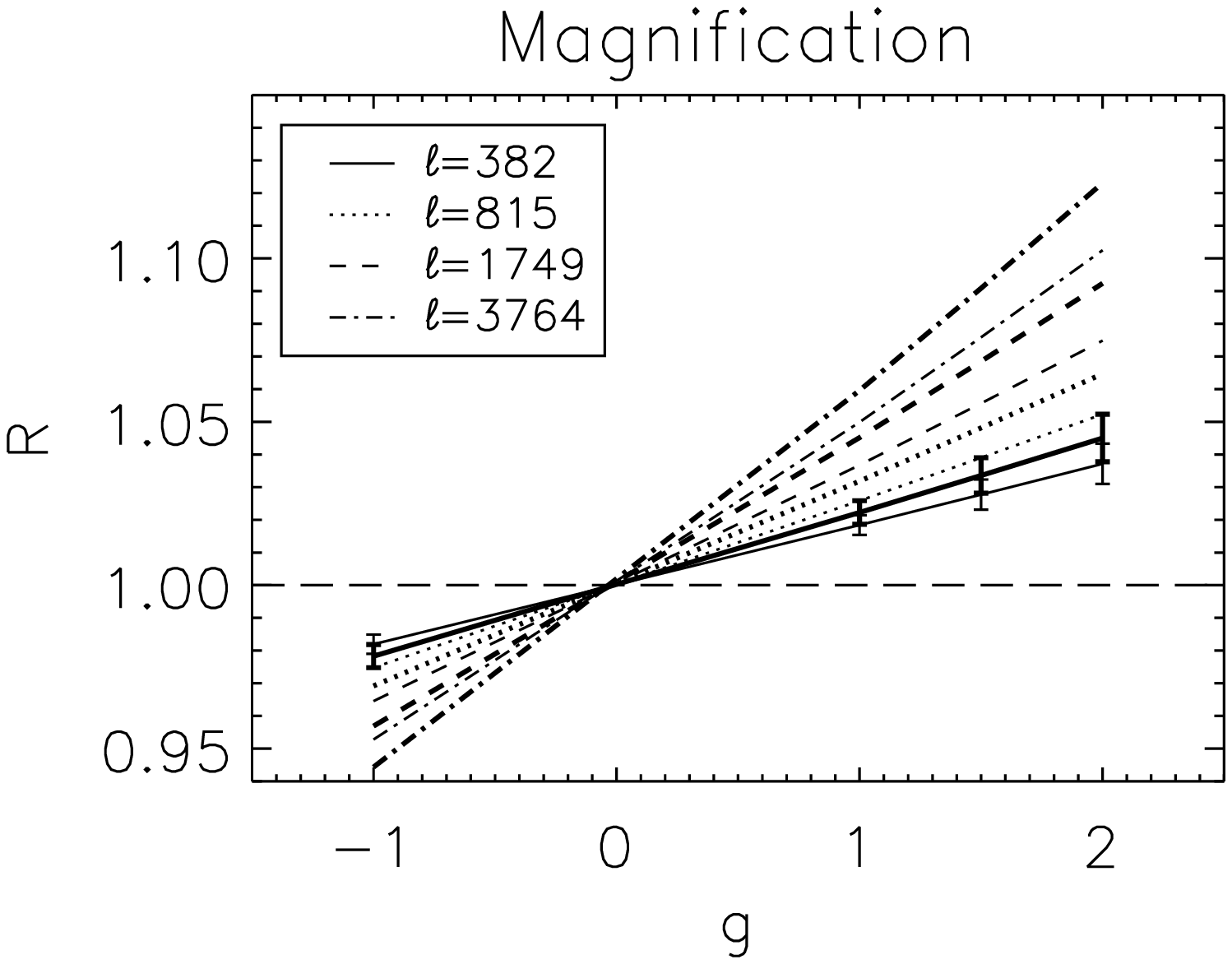}
\caption{The dependence of $R=C_E^{\rm SL}(\ell)/C_E(\ell)$ on $g$ 
for standard estimator.
Results are for photo-z bins $(0.5,0.7)$ (thin lines) and $(0.9,1.1)$ (thick lines).
The data points have $g=-1,1,1.5,2$, respectively.
For clarity, we only plot the error bars for $\ell=382$, which are the rms dispersion among 20 realizations.  
The dependence on $g$ is perfectly linear.
\label{fig:gdgg}} 
\efi
\subsubsection{Dependence on $g$}

Eq. \ref{eqn:CM1} predicts, for the auto-power spectrum, 
\be
\label{eqn:CM2}
R(\ell)-1=c(\ell)g\ .
\ee
Here $c(\ell)$ is determined by the ratio of lensing bispectrum and
power spectrum,  so it  only depends on $\ell$.
This relation is confirmed to high accuracy in Fig. \ref{fig:gdgg},
where results for arbitrarily selected $g=-1,1,1.5,2$ are
presented. This linear relation holds for the two redshift bins and
angular scales investigated ($\ell=382,815,1749,3764$). Notice that
$g$ could be zero or negative,  
leading to no correlation or anti-correlation between the source and lens.  

For the cross correlation, we can derive a similar relation,
\be
R(\ell)-1=c_f(\ell)g_f+c_b(\ell)g_b\ .
\ee
Here, $g_f$ and $g_b$ are the $g$ prefactors of foreground and background galaxies. 
Again, $c_f(\ell)$ and $c_b(\ell)$ are determined by the ratio of 
lensing bispectrum and power spectrum for foreground and background, respectively.
Note that $g$ is also an observable in lensing survey.  
Thus if we further assign the source galaxies into different $g$-bins, 
according to the perfect scaling we could calibrate the SLC effect out.  
Furthermore, this calibration allow us to measure 3rd-order statistics
($\langle\gamma\gamma'\kappa'\rangle$ and $\langle\gamma\kappa\gamma'\rangle$) 
by comparing the power suffering SLC effect with the one calibrated.  

\subsection{Summary}
From the results above, we summarize that the standard estimator
suffers SLC effect induced by both the intrinsic source clustering and the cosmic magnification.
Both of them cause $\mathcal{O}(1\%)$-$\mathcal{O}(10\%)$ overestimation in lensing power spectrum
at angular scale and redshift of interest.
The standard estimator also results in large cosmic variance in the power spectrum measurement.
Such systematic error is significant for lensing cosmology.
Fortunately, SLC induced by the cosmic magnification present a perfect linear dependence on observable $g$.
So it is very promising to self-calibrate it using this linear relation.
SLC induced by the intrinsic source clustering shows nonlinear dependence on $b_g$.
However, in principle, through simulation we could obtain a useful fitting formula and self-calibrate it.


\begin{figure*}
\epsfxsize=8cm
\epsffile{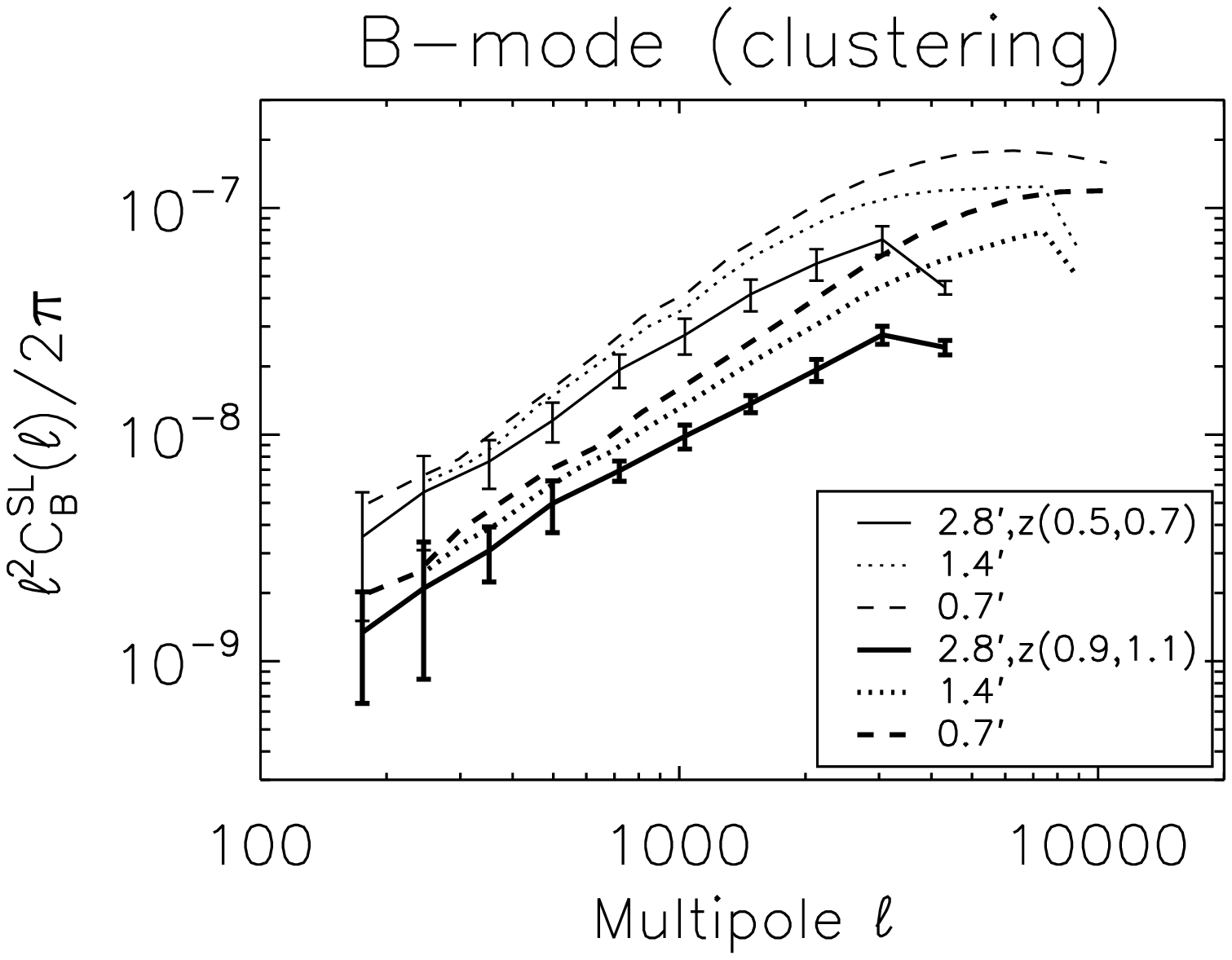}
\epsfxsize=8cm
\epsffile{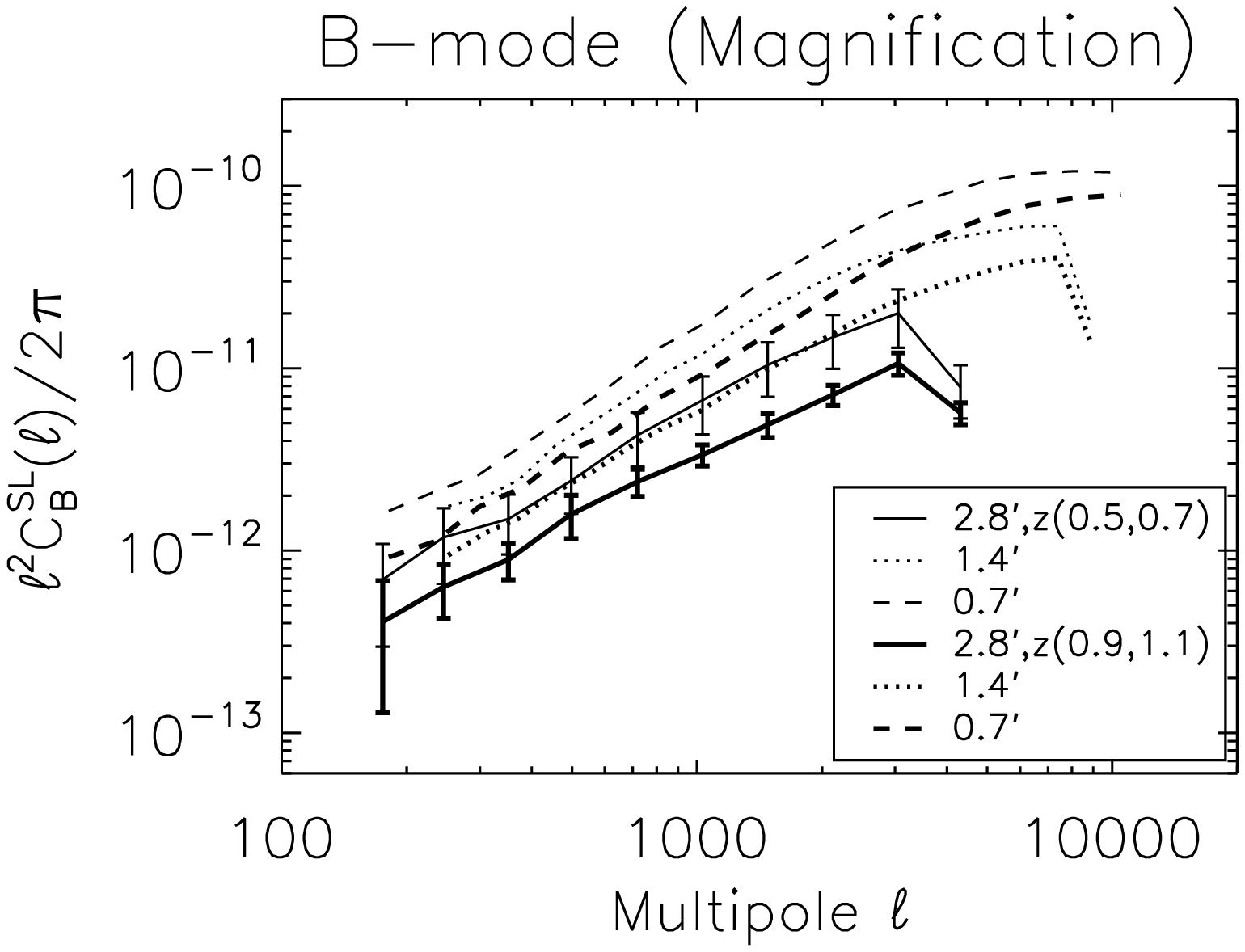}
\caption{The intrinsic source clustering (left panel) and cosmic magnification (right panel) 
induced B-mode power spectrum measured by the pixel-based estimator
(with $b_g=1$ and $g=1$),  for photo-z bins $(0.5,0.7)$ (thin lines)
and $(0.9,1.1)$ (thick lines).  3 pixel sizes, $2.8'$, $1.4'$ and
$0.7'$, are adopted. For clarity, we only plot the error bars for the pixel size of $2.8'$.
The B-mode is robustly identified in all cases and the dependence on pixel size also shows up.
\label{fig:pbbmode}}
\end{figure*}

\begin{figure*}
\epsfxsize=16cm
\epsffile{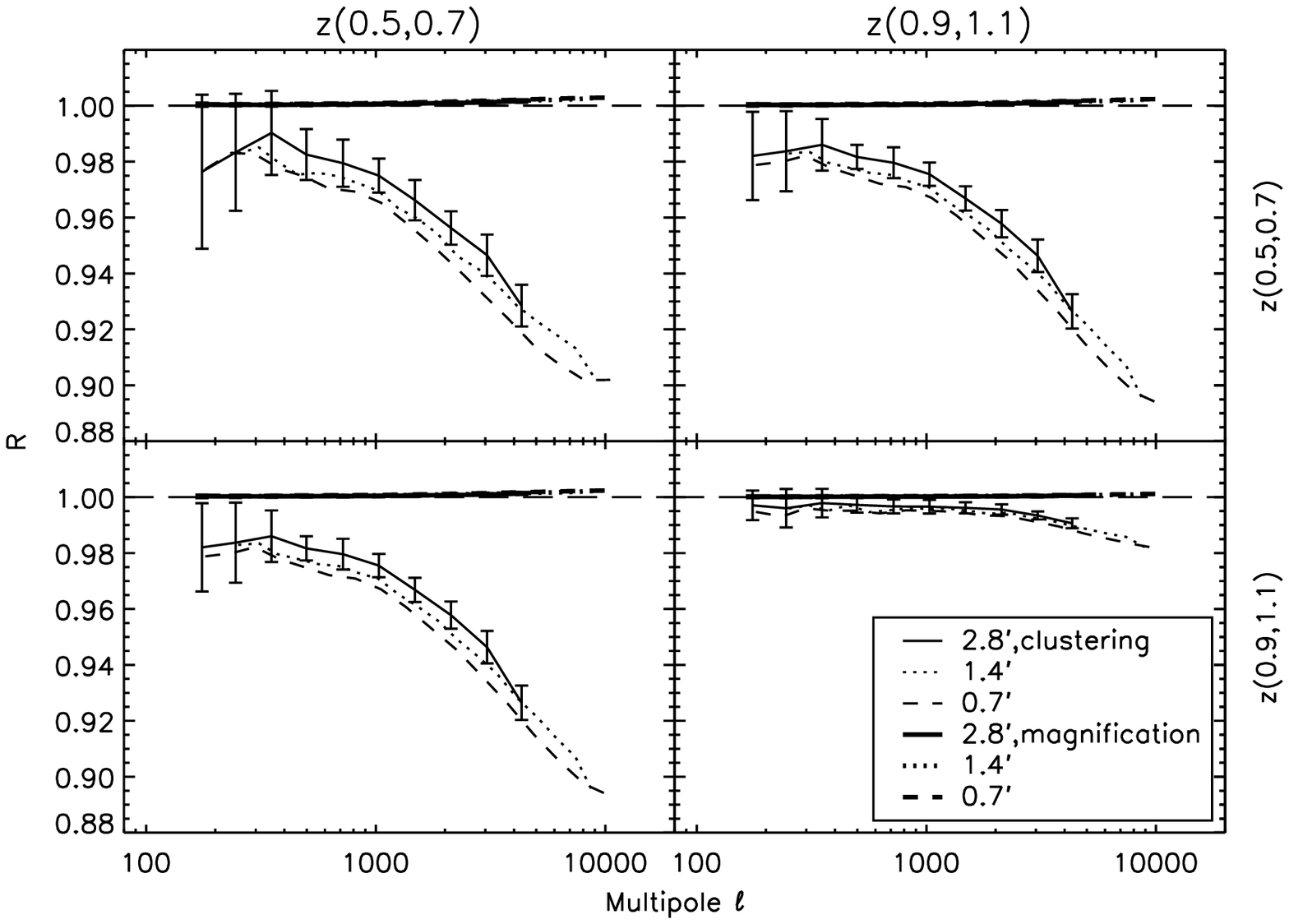}
\caption{The impact of SLC on the lensing power spectrum measured by
  the pixel-based estimator. We plot the ratio $R= C_{E}^{\rm
    SL}(\ell)/C_{E}(\ell)$, for different source redshift, different
  pixel size and different source of SLC.  The pixel-based estimator
  overcorrects the SLC induced by the intrinsic source clustering, causing
  underestimation of the lensing power spectrum by $\sim 1\%$ at
  $\ell=1000$ and $z_s\sim 1$. In contrast, it almost perfectly
  corrects the SLC induced by the cosmic magnification, leading to
  negligible bias in the measured lensing power spectrum. 
\label{fig:pbcps}}
\end{figure*}

\section{SLC with the Pixel-based Estimator}
\label{sec:result2}
This section presents our results on the impact of  SLC on lensing power spectrum
measured with the pixel-based estimator. This estimator first obtains
the averaged cosmic shear on each pixel of the sky and then uses this
map to measure lensing statistics. 

\subsection{Intrinsic Source Clustering}
\label{sec:clustering2}

We take $\delta^\obs=\delta_g$ in Eq. \ref{eqn:pixelbased} to quantify the SLC effect 
due to the intrinsic source clustering, and look at $b_g=1$ first. In
general, all the results in this section depend on the pixel size. We
present results with pixel size $0.7^{'}$, $1.4^{'}$ and $2.8^{'}$ to
show this dependence. But for clarity, we only plot error bars for the
$2.8'$ pixel size case.   
 
\subsubsection{B-mode power spectrum}
The B-mode power spectrum is presented in the left panel of
Fig. \ref{fig:pbbmode}.  A non-zero B-mode is detected at high
significance. Furthermore, it is orders of magnitude higher than the
case without SLC. Hence the B-mode caused by the intrinsic source clustering
induced SLC is robustly identified. It is slightly higher for lower photo-z bin $(0.5,0.7)$, 
which is consistent with the E-mode result below.  This should be due
to stronger intrinsic clustering at lower redshift. 

We also find smaller SLC effect for larger pixel size. Larger pixel
suppresses the intrinsic clustering, leading to smaller SLC. 

\subsubsection{E-mode power spectrum}
The impact of SLC on the lensing power spectrum is shown in Fig. \ref{fig:pbcps}.  
An immediate finding is that it suppresses the E-mode power spectrum
estimation.  
Like the explanation for reduced skewness in \cite{Hamana02},
 this behavior could be understood by considering two patterns.
One pattern is that there is a source over-density near the close-side of the photo-z bin.
Another is that there is an under-density near the close-side.
Averaged over this photo-z bin, the lensing signal is larger in the former pattern.
However, the source over-density close to us lowers the effective distance of the sources, which dilutes the lensing signal.
While the under-density close to us increases the effective distance, leading to increasing of the lensing signal.
This opposite direction of the SLC effect leading to the suppression of 2nd-order statistics. 
At $\ell\sim 1000$ of
particular interest in weak lensing cosmology,  this underestimation
reaches 3\% for the photo-z bin $(0.5,0.7)$ and 1\% for the photo-z bin
$(0.9,1.1)$. It increases towards small angular scale, where the
intrinsic clustering is larger. This underestimation depends on the
pixel size, except at scales much larger than the pixel size (e.g. the
degree scale around $\ell=200$). 

A  surprising result is that the cross-spectrum between the two photo-z bins 
is suppressed more than the auto-power spectrum of each bin.  We argue
that the source clustering of the two redshift bins biases the two lensing
fields incoherently. 
Thus it reduces the coherence between the lensing signal at two redshifts 
and leads to larger suppression for cross-spectrum. 
This result is somewhat consistent with the result in
\cite{Valageas14}, which found that  SLC is important when
cross-correlating low and high redshift bin. 

\subsubsection{Dependence on $b_g$}

The dependence of the detected underestimation on $b_g$ is shown in
Fig. \ref{fig:pbbb}. We find  pretty good linearity for the higher photo-z
bin $(0.9,1.1)$,  and reasonably good for the lower photo-z bin $(0.5,0.7)$.
Since our bias model is very simplified,
further study against mock catalogue is required to robustly quantify $R$-$b_g$ relation.

The detected $1\%$ level bias in the measured lensing power spectrum
is significant for precision lensing cosmology.  Further analysis,
including analysis against mock galaxy catalogue and investigations on PDF
and peak statistics \citep{Liu13},  should be carried out to fully
understand the impact of SLC on weak lensing statistics based on the
pixel-based estimator.

\subsection{Cosmic Magnification}
\label{sec:magnification2}

For the cosmic magnification induced SLC, we find that pixel-based estimator suppresses 
SLC to negligible level, both for the lensing B-mode
(right panel of Fig. \ref{fig:pbbmode}) and the E-mode (Fig. \ref{fig:pbcps}). 
This is valid for all the redshifts, pixel sizes and $g$ investigated. The
detected B-mode has a rms $\sim 10^{-5}$, which is orders of magnitude
smaller than statistical fluctuations in cosmic shear measurement of full-sky survey.
SLC affects the E-mode power spectrum by $\sim 0.1\%$ at $\ell \sim 1000$. 
This bias is again significantly smaller than the statistical accuracy of any
realistic lensing surveys. 

Why does the pixel-based estimator work so well? We argue that it is
due to the fact that $\kappa$ varies slowly across the photo-z bin,
for each line of sight. In the limit that it can be well approximated as a constant
along each line of sight, the weighting $(1+g\kappa)$ in the numerator
and denominator cancel out. So the averaged $\gamma$ is area
averaged, exactly what we want. 
\bfi{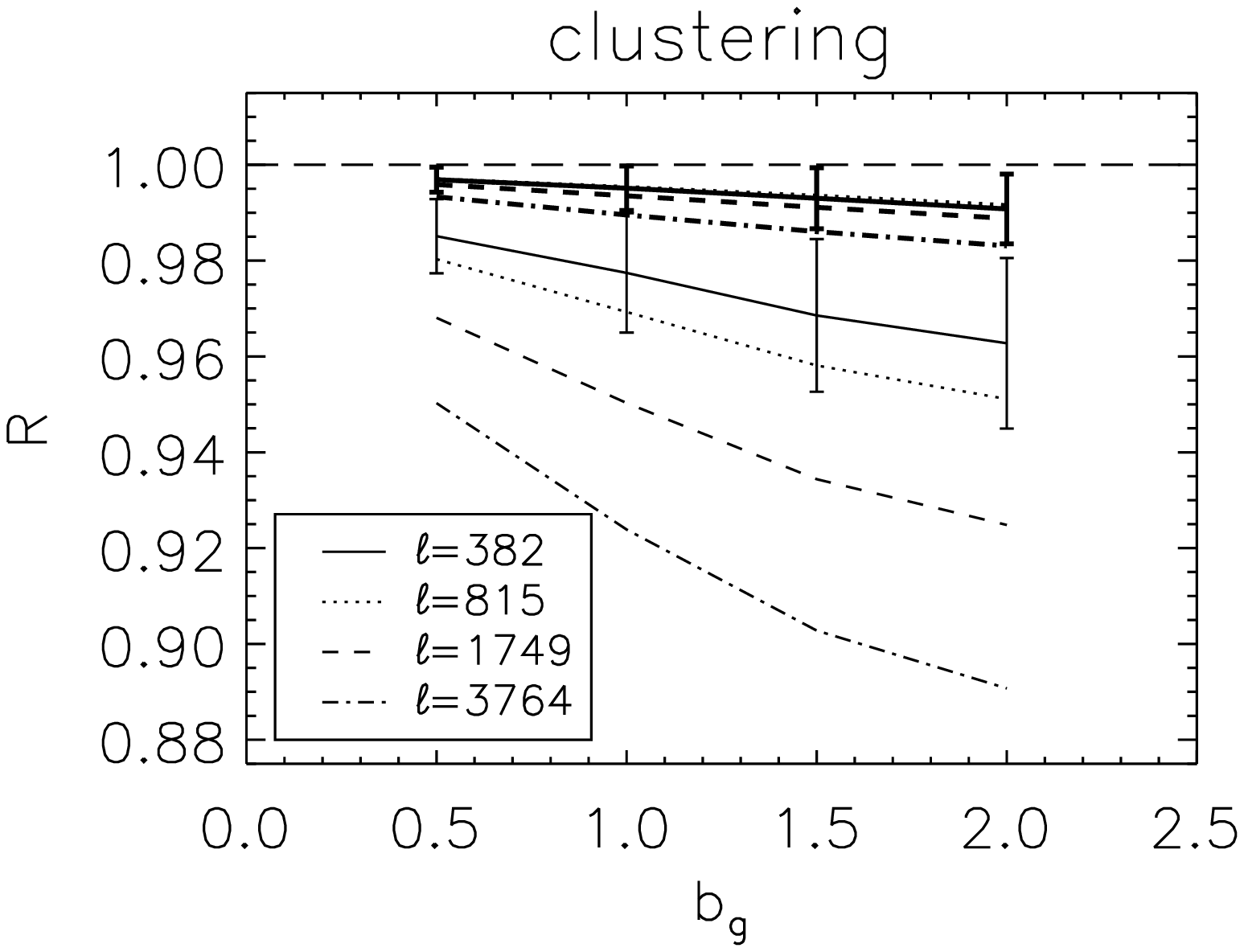}
\caption{The dependence of $R=C_E^{\rm SL}(\ell)/C_E(\ell)$ on the galaxy bias, for the pixel-based
  estimator.
  Results are for photo-z bins $(0.5,0.7)$ (thin lines) and $(0.9,1.1)$ (thick lines).
For clarity, we only plot the errors for $\ell=382$.  
The dependence on galaxy bias shows good linearity.
\label{fig:pbbb}} \efi
\subsection{Summary} 

From the results above, we can conclude  that the pixel-based
estimator is essentially free of SLC induced by the cosmic
magnification. So the SLC effect only arises from the intrinsic source clustering. It
causes $1$-$10\%$ underestimation of the lensing power spectrum at
angular scale and redshift of interest. Such systematic error  is
significant for lensing cosmology. Fortunately we find that the
induced SLC has a reasonably good linear dependence on the galaxy
bias. So  it is promising to self-calibrate the SLC effect using this linear
relation. 


\section{Conclusion and Discussion}
\label{sec:conclusion}

We use N-body simulation to evaluate the SLC effect 
in the context of weak lensing tomography with realistic photo-z errors.  
We quantify the effect for two kinds of estimator, the standard
estimator and the pixel-based estimator,  
and for two sources of the SLC effect, the intrinsic source clustering and
the cosmic magnification.   
The statistics investigated include B-mode power spectrum, the ratio
of E-mode power spectra,  
and the dependence on galaxy bias or magnification prefactor.

We found that in all cases, B-mode is clearly produced by the SLC effect.
But the amplitude of the produced B-mode is at least 3 orders of magnitude lower than the E-mode.
Thus these B-modes are not expected to be detected in near future lensing survey.

For the standard estimator, the SLC effect induced by both the intrinsic source clustering
 and the cosmic magnification can bias the lensing power spectrum by $\mathcal{O}(1\%)$-$\mathcal{O}(10\%)$
 at the scale and redshift of interest.
Furthermore, the standard estimator brings large cosmic variance into the power spectrum measurement.
But the cross power spectrum suffers much less SLC effect induced by the intrinsic source clustering.
These two sources of SLC together make the situation more complicated.  
Fortunately, the SLC effect shows perfect linear dependence on the observable $g$,
 which comes free from the lensing survey.
Thus the self-calibration is very promising.
Large sky coverage will beat down the fluctuations in the SLC effect induced by the intrinsic source clustering.
When this uncertainty is well under control, in principle, we could find a fitting formula for the dependence on galaxy bias
and utilize it to the observation to self-calibrate the SLC effect induced by the intrinsic source clustering.
 
For the pixel-based estimator, we found negligible cosmic magnification induced SLC effect.  
However, the intrinsic source clustering induced SLC suppresses the power spectrum 
measurement by $\mathcal{O}(1\%)$-$\mathcal{O}(10\%)$.
Fortunately, the dependence of SLC effect on galaxy bias shows good linearity.
This linear scaling for low bias galaxies and for high redshift is also useful to self-calibrate the SLC effect. 

From the results in this work, the pixel-based estimator has negligible SLC effect from the cosmic magnification.
The intrinsic source clustering induced SLC effect shows good linear dependence on galaxy bias.
Thus it is easy to deal with SLC effect by self-calibration.
However, the standard estimator suffers sever SLC effect from both the two sources
 and large cosmic variance induced by the intrinsic source clustering.
Although the dependence on $g$ is perfectly linear, we need further works to deal with the SLC effect
 induced by the intrinsic source clustering.
 
For the standard estimator, the B-mode produced by intrinsic source clustering is
consistent with the result in \cite{Schneider02},
in which they analytically estimated the amplitude.  
We caution that lensing tomography indeed could suppress SLC effect induced by intrinsic source clustering,
but not to a negligible level.  
The large SLC effect induced by cosmic magnification is consistent with the result in \cite{Schmidt09}, 
in which they focused on lensing bias and analytically calculated the correction term from bispectrum.  
As pointed out in the literatures, the effect from approximating reduced shear as cosmic shear 
shares the same form with the cosmic magnification ($g=1$).  
Our result is also consistent with the result for reduced shear in \cite{Dodelson06}.
Thus our work here also confirms the result for reduced shear through simulation.  

The SLC effect is a severe systematic error in weak lensing cosmology.
We need further investigations to deal with it. The bias in
cosmological parameters is one important focus. The goodness of the
above self-calibration method is also of great importance. Numerical
results presented in this paper are limited by the finite box size and
moderate resolution and limited realization of the analyzed N-body
simulation. They should be rechecked and improved by larger
simulations with higher resolution and many more
realizations. Furthermore,  we need more realistic lensing mock
catalogue, with various types of galaxies,  
various photometric redshift distributions and bin schemes,  to test these scaling relations and methods.

We emphaszie that the two estimators are not redundant. In contrast,
they are complementary to each other and provide a valuable
cross-check on the lensing measurement and the SLC effect. (1) The standard estimator is convenient for  direct measurement of 2-point statistics from lensing survey.
(2) The pixel-based estimator is of great important for map-making and
higher-order  statistics such as bispectrum and 1-point PDF.
(3) Given that the underlying lensing signal is the same and given
that the two estimators suffer differently from the SLC effect, with
different dependence on galaxy bias and magnification prefactor, 
it is interesting to compare the two independent lensing measurements.
Due to the negligible SLC induced by the cosmic magnification and reasonably good $R$-$b_g$ relation,
 it seems easier to get unbiased lensing signal from the pixel-based estimator.
Combining this unbiased lensing signal with the one derived from the standard estimator, 
it is possible to identify and isolate the SLC effect in observation.

\section*{Acknowledgment} 
This work was supported by the national science foundation of China 
(Grant No. 11025316, 11121062, 11033006, 11320101002, 
10873027，11233005, 11473053, U1331201).
The work is also supported by the ''Strategic Priority Research Program the
Emergence of Cosmological Structures'' of the Chinese Academy of Sciences, Grant No. XDB09010400.
This work made use of the High Performance Computing Resource in the Core Facility for Advanced Research Computing at Shanghai Astronomical Observatory.

\appendix

\section{Simulation}
\label{sec:simulation}

Our N-body simulation was run using the Gadget-2 code \citep{Gadget2}
adopting the standard $\Lambda$CDM cosmology, 
with $\Omega_m=0.266$, $\Omega_{\Lambda}=1-\Omega_m$,
$\sigma_8=0.801$, $h=0.71$ and $n_s=0.963$.
It is performed with $512^3$ particles in the $300\mpch$ box.
The simulation details can be found in \cite{Cui10}. 
For the purpose of lensing study, 
the output redshifts $z_i$ are specified such that any two adjacent snapshots 
are separated by the box size $L=300\mpch$ in comoving distance,
i.e. $D(z_i)=300i-150 \mpch (i=1,2,3, \cdots)$.
For a source redshift $z_s\sim 1$, we need to stack 10 snapshots to
construct one lensing map.   
Since each snapshot is obtained from the same initial conditions,
to avoid artificial correlation, we randomly shift and rotate the boxes utilizing the periodical boundary condition.
All quantities involved in SLC effect (matter overdensity, convergence, cosmic shear) are constructed as follows. \\
(1) We stack 10 randomly shifted and rotated boxes to comoving distance $3000\mpch$. 20 light-cone realizations are made with sky coverage of $6.03^{\circ}\times 6.03^{\circ}$ each.  \\
(2) We cut the light-cones into density slices $\delta_{m,i}$ $(i=1,\cdots,140)$ with width $\Delta z=0.01$, 
from $z=0$ to $z=1.4$, adopting NGP mass-assignment.\\ 
(3) Adopting Born approximation, we construct $\kappa_i$ slices also with $\Delta z_s=0.01$, according to 
\be
\label{eqn:kappa}
\kappa(\hat{n},z_s)=\int_0^{\tilde\chi(z_s)}\delta_m(\hat{n},z)W(z,z_s)\dif\tilde{\chi}
\ee
and the corresponding discrete form
\be
\label{eqn:kappaD}
\kappa_i(\hat{n})=\sum_{j=1}^{i-1}\delta_{m,j}(\hat{n})W_{ji} \Delta z\ .
\ee
Here, $\delta_m$ is the matter overdensity.  
$\chi\equiv \chi(z)$ is the comoving angular diameter distance to the lens redshift $z$.
We conveniently express $\chi$ in units of the Hubble radius, $\tilde{\chi}\equiv \chi/(c/H_0)$,
in which $H_0$ is the Hubble constant today.  
The lensing kernel $W(z,z_s)$ for a source at redshift $z_s$ and a lens at redshift $z$ is given by
\be
W(z,z_s)=\frac{3}{2}\Omega_m (1+z)\tilde{\chi}(z)\left[1-\frac{\chi(z)}{\chi(z_s)}\right] 
\ee
when $z\leq z_s$ and zero otherwise.
The corresponding discrete form is
\be
W_{ji}=\frac{3}{2}\Omega_m(1+z_j) \tilde{\chi_j}\left[1-\frac{\chi_j}{\chi_i}\right]\ .
\ee
Here $\Omega_m$ is the cosmological matter density of the universe in units of the critical density.
The above expression is valid for the flat cosmology we consider throughout the paper.\\ 
(4) We convert $\kappa$ slices into cosmic shear slices through the relation in Fourier space 
\be
\gamma_1(\vec\ell)=\kappa(\vec\ell)\cos2\varphi_{\vec\ell}\ ,\hspace{5mm}
\gamma_2(\vec\ell)=\kappa(\vec\ell)\sin2\varphi_{\vec\ell}\ .
\ee\\
(5) We treat the source galaxy distribution as the number density field $\delta^\obs(\hat{n},z)$. 
For the investigation on the intrinsic source clustering, we adopt a simple bias model $\delta^\obs(\hat{n}, z)=b_g\delta_m(\hat{n},z)$.
While for the cosmic magnification, only leading term is kept, i.e. $\delta^\obs(\hat{n}, z)=g\kappa(\hat{n}, z)$.
For general study we direct adopt $b_g=1$ and $g=1$.
For the dependence study on $b_g$ and $g$, we just multiply several typical values to construct $\delta^\obs(\hat{n}, z)$ for each cases.
We choose $b_g=0.5, 1, 1.5, 2.0$ and $g=-1, 1, 1.5, 2$.
To avoid unrealistic $\delta^\obs(\hat{n}, z)<-1$, which could happen if $b_g>1$ or $g\gg 1$,
we simply set $\delta^\obs(\hat{n}, z)=-1$ where it happens. We then
integrate over the corresponding source redshift distribution to
obtain the projected galaxy number overdensity and lensing signal.

\clearpage

\end{document}